\documentclass[reqno,letterpaper,12pt]{article}
\usepackage{amsmath,amsthm,amscd,amsfonts,amsxtra,amssymb}
\usepackage{graphicx}
\usepackage{psfrag}
\usepackage{marvosym}
\usepackage{pstricks}

\numberwithin{equation}{section}
\numberwithin{thm}{section}
\numberwithin{figure}{section}

\newcommand{\extraspace}{\addtolength{\abovedisplayskip}{2mm}
                        \addtolength{\belowdisplayskip}{2mm}
                        \addtolength{\abovedisplayshortskip}{2mm}
                        \addtolength{\belowdisplayshortskip}{2mm}}
\newcommand{\be}{\begin{equation}\extraspace}
\newcommand{\ee}{\end{equation}}
\newcommand{\bea}{\begin{eqnarray}\extraspace}
\newcommand{\eea}{\end{eqnarray}}
\newcommand{\bsp}{\begin{split}}
\newcommand{\esp}{\end{split}}

\newcommand{\wtPsi}{\widetilde{\Psi}}

\newcommand{\nonu}{\nonumber \\[2mm]}

\newcommand{\ziu}{z_i^\uparrow}
\newcommand{\zju}{z_j^\uparrow}
\newcommand{\zku}{z_k^\uparrow}
\newcommand{\zoneu}{z_1^\uparrow}
\newcommand{\ztwou}{z_2^\uparrow}
\newcommand{\zid}{z_{i'}^\downarrow}
\newcommand{\zjd}{z_{j'}^\downarrow}
\newcommand{\zld}{z_{l'}^\downarrow}
\newcommand{\zoned}{z_{1'}^\downarrow}
\newcommand{\ztwod}{z_{2'}^\downarrow}

\newcommand{\woneu}{w_1^\uparrow}
\newcommand{\wtwou}{w_2^\uparrow}

\newcommand{\su}{\sigma_\uparrow}
\newcommand{\sd}{\sigma_\downarrow}
\newcommand{\sthree}{\sigma_3}
\newcommand{\up}{\uparrow}
\newcommand{\down}{\downarrow}

\newcommand{\cF}{{\mathcal F}}

\newcommand{\cN}{{\mathcal N}}

\newcommand{\NN}{{\mathbb N}}

\newcommand{\ZZ}{{\mathbb Z}}
\newcommand{\id}{{\mathbf 1}}

\newcommand{\qnum}[1]{\lfloor #1 \rfloor}

\newcommand{\ua}{\uparrow}
\newcommand{\da}{\downarrow}

\newcommand{\vep}{\varepsilon}

\DeclareMathOperator{\mmod}{mod}

\newcommand{\beq}{\begin{equation}}
\newcommand{\eeq}{\end{equation}}

\def\dottedhline(#1,#2)#3#4#5{%
\multiput(#1,#2)(#3,0){#4}{\circle*{#5}}
}

\newlength{\strutlengte}
\newlength{\strutdiepte}
\newlength{\strutbreedte} 
\newlength{\extraonderruimte}
\newlength{\extrabovenruimte}
\allowdisplaybreaks[4]

\usepackage{graphicx}
\begin{document}

\begin{center}
{\Large\bf Wavefunctions for topological quantum registers}\\[1cm]

{\large E. Ardonne} \\[1mm]
{\it Microsoft station Q, Kavli Institute for Theoretical Physics\\
University of California, Santa Barbara, CA 93106, USA and\\
Center for the Physics of Information, \\
California Institute of Technology, Pasadena, CA 91125, USA}\\[4mm]
             
{\large K. Schoutens} \\[1mm]
{\it	Institute for Theoretical Physics, Valckenierstraat 65 \\
	1018 XE  Amsterdam, the Netherlands}\\[2mm]

June 8, 2006
 
\end{center}

\begin{quotation}{\small\noindent
We present explicit wavefunctions for quasi-hole excitations over 
a variety of non-abelian quantum Hall states: the Read-Rezayi states 
with $k\geq 3$ clustering properties and a paired spin-singlet quantum 
Hall state. Quasi-holes over these states constitute a topological 
quantum register, which can be addressed by braiding quasi-holes.
We obtain the braid properties by direct inspection of the quasi-hole
wavefunctions. We establish that the braid properties for the
paired spin-singlet state are those of `Fibonacci anyons', and 
thus suitable for universal quantum computation. Our derivations 
in this paper rely on explicit computations in the parafermionic 
Conformal Field Theories that underly these particular quantum Hall 
states.}
\end{quotation}

\section{Introduction}

The realization that quantum Hall systems may harness what is
called non-abelian braid statistics has led to two exciting prospects.
The first is that experiments can be set up where, for the first time,
the existence of non-abelian statistics in nature can be established.
The second prospect, following early ideas of Kitaev \cite{kit1}, is that
systems exhibiting non-abelian braid statistics can give rise to 
quantum registers and may offer unique possibilities for what has
come to be known as topological quantum computation or fault 
tolerant quantum computation.

Proposals for experimental detection of non-abelian statistics 
\cite{fntw,dfm,sh1,bks} have
mostly focused on the so-called $\nu=5/2$ quantum Hall state, which
is believed to be described by the `pfaffian' or Moore-Read state 
\cite{mr1,morf}.
To this date, this is the single quantum Hall plateau where the 
indications for an underlying non-abelian state are rather firm. 
For the plateau at 
$\nu=12/5$ there is a more speculative case for a connection with a 
non-abelian $k=3$ Read-Rezayi state \cite{rr2}.
Proposals to test if the
state observed at $\nu=12/5$ is indeed non-abelian have been
described in \cite{bss,cs1}.

An interesting domain for the search for non-abelian quantum Hall states 
is that of multi-component quantum Hall states, which can be realized 
as double layer or as spin-singlet states. The double layer case in 
particular may offer the experimental flexibility needed for tuning into
a regime where a non-abelian state takes the upper hand. Examples of 
candidates are the spin-singlet (or double layer) states of \cite{alls}, 
exhibiting a separation of spin and charge, and the spin-singlet analogues 
of the Moore-Read and Read-Rezayi states that we introduced in \cite{as1}
and studied further (with N.\ Read and E.\ Rezayi) in \cite{arrs}.

Ideas for topological quantum computation in quantum Hall systems boil down
to the following. Having available a non-abelian quantum Hall state as the
underlying medium, the injection of quasi-hole excitations is known to open up
an internal space (the `quantum register'), whose dimensionality grows with
the number of quasi-holes. This quantum register can be addressed by performing
adiabatic quasi-hole braidings, which give rise to matrices acting on the 
register.

It is well-known that the non-abelian braiding in the Moore-Read state is
not sufficiently rich to enable universal quantum gates on the quantum
register (note that there are proposals to combine topological
and non-topological operations to obtain universal gates in this state
\cite{bk1,bravyi,fnw}). It is also known that `higher' non-abelian quantum 
Hall states, such as those in the Read-Rezayi series with $k=3,5,6,\ldots$, 
do offer a prospect of universal quantum computation \cite{flw1,flw2}, in the 
sense that universal quantum gates, such as the 2-qubit CNOT gate, can be 
approximated with arbitrary precision by a well-chosen sequence of braid 
matrices \cite{solkit,bhzs}. In this paper, we shall establish that the
paired spin-singlet state (AS state) of \cite{as1} is also universal for 
quantum computation. It is the unique paired quantum Hall state with this 
property. 

The simplest scenario for braid matrices with sufficient structure for 
universal topological quantum computation is offered by the so-called 
`Fibonacci anyons'. {\it In abstracto}, these
are particles of two types, `0' and `1', with fusion rules
\be
  0 \times 0 = 0 \ , \qquad   0 \times 1 = 1 \ , \qquad   1 \times 1 = 0+1 \ . 
\label{eq:fibofusion}
\ee
Through a universal connection between, on the one hand, fusion rules 
and, on the other, braid matrices, it has been established that the 
simplest braid matrices for Fibonacci anyons are, in the notation of 
\cite{pr},
\bea
&&
R= \left( \begin{array}{cc}
              (-1)^{4/5} & 0 \\
              0 & (-1)^{-3/5} 
           \end{array} \right) \, \quad 
F= \left( \begin{array}{cc}
             \tau & \sqrt{\tau} \\
             \sqrt{\tau} & - \tau 
              \end{array} \right) \, 
\nonu
&& \qquad 
B= (-1)^{-4/5}\left( \begin{array}{cc}
             \tau & (-1)^{-3/5} \sqrt{\tau} \\
              (-1)^{-3/5} \sqrt{\tau} & (-1)^{-1/5} \tau 
              \end{array} \right) \ ,  
\label{eq:RFB}
\eea
where $\tau=\frac{1}{2}(\sqrt{5}-1)$.

Quasi-holes over non-abelian quantum Hall states cannot straightforwardly be 
identified with `non-abelian anyons', but their are important parallels, 
in particular where the fundamental relations between fusion and braiding 
properties are concerned. These relations were first studied in the context 
of the algebraic approach rational conformal 
field theories, as developed by Moore and Seiberg in \cite{ms1}. In the 
context of quantum 
Hall systems, these same relations have been exemplified in the work of 
Slingerland and
Bais \cite{sb1}, who used an associated quantum group structure to obtain 
explicit results for braid 
matrices for the Read-Rezayi quantum Hall states for general $k$.

Before the work of \cite{sb1}, Nayak and Wilczek \cite{nw1} had derived 
explicit wavefunctions for four quasi-hole excitations over the Moore-Read 
state. These wavefunctions provide detailed information on the internal 
state associated with four quasi-holes, as a function of the 
locations of these excitations. {}From these wavefunctions, braid properties 
are derived by direct inspection. The work of \cite{nw1} was based on a 
`coordinate' rather than 
an `algebraic' approach to (rational) Conformal Field Theory (CFT), and 
employed bosonization
techniques for mastering the relevant CFT correlators.

It is the purpose of this paper to present explicit expressions for 
wavefunctions of quantum
registers associated to non-abelian quantum Hall states that are sufficiently 
rich to enable universal 
topological quantum computation. We will in particular focus on two distinct 
quantum Hall states
that both give rise to braid matrices of the type displayed in 
eq.~(\ref{eq:RFB}) (up to additional abelian phase factors). The first is the 
$k=3$ Read-Rezayi state and the second is the paired ($k=2$) AS spin-singlet 
quantum Hall state. We shall also write some of the wavefunctions for 
quasi-holes over the general $k$ Read-Rezayi states. Note that throughout this paper,
we will assume that the only effect of braiding comes from the explicit monodromy.

The appearance of the `Fibonacci-type' braid matrices in the quantum Hall 
systems can be understood from a coarse-graining of the fusion rules of 
the parafermionic CFTs underlying 
these states. For the example of the spin-singlet state, this takes the form
\be
\text{`$0$'} = \{ \id, \psi_1, \psi_2, \psi_3 \} \ , \qquad 
\text{`$1$'} = \{ \sigma_1, \sigma_2, \sigma_3, \rho\} 
\ee
with the $\psi_i$ denoting the parafermion sectors and $\sigma_i$, $\rho$ 
labeling the
various parafermion spin fields in the CFT. The fusion rules of these fields 
(see table \ref{fusrul} in section \ref{sec:su32pf})
are such that the coarse graining into `0' and `1' respects the relations 
given in eq.~(\ref{eq:fibofusion}).
We would like to stress that the `Fibonacci anyon' aspect of the quantum Hall 
quasi-holes 
captures a limited fraction of their relevant properties: for example, the 
fundamental quasi-holes 
come with different sets of quantum numbers and the detailed fusion rules (and 
Operator
Product Expansions) of the fields in the parafermionic CFTs lead to detailed 
structure
expressed in the wavefunctions that we derive in this paper.

The `coordinate CFT' approach that we follow to derive the quasi-hole 
wavefunctions 
delves deep into the results for Wess-Zumino-Witten (WZW) and parafermion 
CFTs derived in the mid-1980s. We shall in particular rely on results of 
Knizhnik and Zamolodchikov (KZ) \cite{kz1}. Their expression for 4-point 
functions of particular primary 
fields in WZW models will form the cornerstone of the `contraction arguments' 
that we employ to determine closed form expressions for the various quantum 
register states that we consider. The `master formulas' that we develop 
enable an easy evaluation of correlators that have until now not appeared 
in the literature, and that are not easily computed with the methods of KZ.
We present some explicit examples in sections \ref{sec:morecorrsu23},
\ref{sec:morecorrsu32}.

The presentation in this paper is organized as follows. In section \ref{genapp},
we will explain the method we use to obtain the quasi-hole correlators by using
the Moore-Read state as an example. In section \ref{rrsec}, we apply this
method to the $k=3$ Read-Rezayi states. We will provide a detailed derivation
of the quasi-hole wavefunctions, and use these to calculate the braid behaviour
of the quasi-holes. In section \ref{sec:asstates}, we apply our method to the
paired spin-singlet states proposed by the authors. In section \ref{sec:qg} we
compare the braid results of sections \ref{rrsec} and \ref{sec:asstates} with the
results obtained by using the associated quantum groups.
In the two appendices \ref{su2app} and \ref{su3app}, we give the details of the parafermion
CFT's corresponding to $su(2)_k$ and $su(3)_2$ respectively. Namely, we
provide the fusion rules, the details of the Operator Product Expansions,
including the various coefficients, the spin-field correlators used to derive the
quasi-hole wavefunctions and the braid behaviour of the various hypergeometric
functions which show up in the correlators.  In addition, we give some new
parafermionic correlators, which were used to obtain the various OPE
coefficients. In section \ref{sec:rrk}, we give the quasi-hole wavefunctions
for the Read-Rezayi states for arbitrary $k$. 

\vskip 8mm

\noindent
{\em Acknowledgments} EA thanks J.K.\ Slingerland, M.\ Stone and Z.\ Wang for many
valuable discussions; KS acknowledges discussions with A.N.\ Schellekens
and E.P.\ Verlinde and thanks the KITP at Santa Barbara for hospitality
while this work was done. This research was supported in part by the 
National Science Foundation under Grant No.\ PHY99-07949.

\section{General form of quasi-hole wavefunctions}
\label{genapp}

In our analysis of non-abelian quantum Hall states, we rely on the so-called
qH-CFT connection where wavefunctions of a quantum Hall (qH) system are 
expressed as chiral correlators (conformal blocks) of an associated Conformal 
Field Theory (CFT) \cite{mr1}. The connection hinges on the identification in 
the CFT of an {\it electron operator}\ $\psi_{\rm e}(z)$, carrying charge 
$q=-1$. The quantum Hall wavefunction is then expressed as 
\be 
\Psi_{\rm qH} (z_1,\ldots,z_N) =
\lim_{z_\infty \to \infty}
(z_\infty)^\frac{N^2}{\nu} 
\left<
\psi_{\rm e}(z_1)\psi_{\rm e}(z_2) \cdots \psi_{\rm e} (z_N) 
Q_{\rm bg}(z_\infty)
\right>
\ee
with $Q_{\rm bg}$ representing a neutralizing background (ionic) charge
and the factor $(z_\infty)^\frac{N^2}{\nu}$ is included to obtain a
non-vanishing result. (The case of spin-full fermions has some additional 
structure.) Note that we drop the Gaussian factors throughout this paper.

The injection of a quasi-hole at position $w$ is represented by the insertion 
in the CFT correlator of the {\it quasi-hole operator}\ $\phi_{\rm qh}(w)$. 
Mutual locality of the quasi-holes and the electrons in the quantum Hall 
condensate 
implies that the Operator Product Expansion (OPE) between electron and 
quasi-hole operators is of the 
form
\be
\phi_{\rm qh}(w) \psi_{\rm e}(z) = (w-z)^n \phi'_{\rm qh} (z)
\ee
with $n$ a non-negative integer. Note that the mutual locality puts a constraint
on the possible quasi-hole operators.

In all cases studied in this paper, the electron and quasi-hole operators
are expressed in terms of free bosonic fields (representing charge and spin) 
and of a parafermionic CFT. The latter are closely related to 
Wess-Zumino-Witten (WZW) theories. For the order-$k$ Read-Rezayi (RR) states 
the `CFT-data' are: $SU(2)_k$ WZW theory and the associated $\ZZ_k$ 
parafermions, while the paired spin-singlet states are connected to the  
$SU(3)_2$ WZW theory and to the associated higher-rank parafermions. 
We refer to \cite{thesisA} for a review and further details.

For the Moore-Read (MR) state (which is the $k=2$ member of the RR series), 
the parafermion 
theory reduces to a single real (Majorana) fermion $\psi(z)$. This allows a 
direct evaluation of the electron wavefunction using the Wick theorem, 
leading to a `pfaffian' wavefunction. Quasi-holes over this pfaffian state 
can be characterized by pair braking in the pfaffian BCS factor \cite{rr1}. 
The full dependence of a multi-quasi-hole wavefunction on all coordinates is 
set by a CFT correlator
\be
\left<
\sigma(w_1) \cdots \sigma(w_n)
\psi(z_1) \cdots\psi(z_N)
\right>^{(0,1)} 
\ee
with $\sigma(w)$ the spin-fields for the Majorana fermion. In a 1996 
paper, Nayak and Wilczek used bosonization techniques to derive an explicit 
expression for the full four quasi-hole wavefunction \cite{nw1}. It has the general form
\bea
\label{eq:MR4qh}
\lefteqn{
\Psi^{(0,1)}_{\rm MR} (w_1,w_2,w_3,w_4;z_1,z_2,\ldots,z_N) =}
\nonu
&&
A^{(0,1)}(\{w\}) \Psi_{12,34} (\{w\},\{z\}) + B^{(0,1)}(\{w\})  
\Psi_{13,24}(\{w\},\{z\})  \ .
\eea
In this expression, the factors $\Psi_{12,34}$ and $\Psi_{13,24}$,
which are polynomial in all coordinates $(\{w\},\{z\})$, represent independent
(in this case: two) ways in which four quasi-holes can break up pairs in the
electron condensate. The prefactors $A^{(0,1)}$ and $B^{(0,1)}$ are given by
(note that we are giving the result for the fermionic case $M=1$)
\begin{align}
A^{(p)} (\{w_i\}) = & \frac{(-1)^{-\frac{p}{2}}}{2} (w_{12}w_{34})^\frac{1}{4} x^{\frac{1}{4}}
\left( (-1)^p \sqrt{1-\sqrt{x}}+ \sqrt{1+\sqrt{x}}\right) 
\nonu
B^{(p)} (\{w_i\}) = & -\frac{(-1)^{-\frac{p}{2}}}{2} (w_{12}w_{34})^\frac{1}{4}
x^{-\frac{1}{4}} (1-x)^{\frac{1}{2}}
\left(-\sqrt{1-\sqrt{x}} +(-1)^p \sqrt{1+\sqrt{x}}\right) \ , 
\end{align}
where $p=0,1$ and we used the anharmonic ratio
$x=\tfrac{w_{12}w_{34}}{w_{14}w_{32}}$,
(where $w_{ij} = (w_i-w_j)$), which is the same as the anharmonic ratio 
used in \cite{kz1}, but differs from the one used in \cite{nw1}.
The labels $(0,1)$ refer to the fusion channel of the 4-quasi-hole state; it is
this index which acts as the qubit index in the context of topological
quantum computation. 

The formula (\ref{eq:MR4qh}) admits an elegant interpretation: it describes 
precisely how the fusion channel basis of the 4-quasi-hole states decomposes 
over a basis set by patterns in quasi-hole induced pair-breaking in the 
condensate. This decomposition is given as a function of the quasi-hole 
coordinates $w_i$, meaning that it can be followed as quasi-particles 
move through the 
condensate. This in particular implies that quasi-hole braiding properties 
can be read off from these wavefunctions, as was done in \cite{nw1}. 
Clearly, the information stored in these wavefunctions goes well beyond
braiding properties; we expect that some of this additional structure will 
be relevant for the optimal design of experimental protocols aimed at 
demonstrating non-abelian statistics and at quantum computation. 

In this paper we show that quasi-hole wavefunction for non-abelian
quantum Hall  states with potential for universal topological quantum 
computation can be cast in a form similar to (\ref{eq:MR4qh}).
To achieve this goal, we rely, in a first step, on known expressions for 
the multi-parafermion correlators representing the quantum Hall condensate 
in the 
absence of excitations (\cite{rr2,cgt1,arrs,sal1}). Analyzing the factors 
associated
with the injection of quasi-holes (of various kinds) then leads to 
`master formulas' not unlike (\ref{eq:MR4qh}). In a final step we consider 
various `contractions' of this master formula and use those to relate the 
coefficients such as $ A^{(0,1)}$ and $ B^{(0,1)}$ to correlators having 
just four parafermion spin-fields. The latter can extracted from \cite{kz1}, 
where they have been expressed in terms of hypergeometric functions.  

\section{Quasi-hole wavefunctions for the $k=3$ Read-Rezayi states}
\label{rrsec}

In this section, we use the approach outlined in the previous section
to obtain the wavefunctions of quasi-holes over the $k=3$ Read-Rezayi (RR) 
states, \cite{rr2}, which can be viewed as clustered analogues of the paired 
Moore-Read (MR) state \cite{mr1}.
The RR states have been studied in detail by various authors. Among the 
advances are explicit wavefunctions in terms of the electron coordinates, 
both with and without the presence of quasiholes. 

Even though the structure of the quasi-hole wavefunctions in terms of the 
electron coordinates is known, see \cite{cgt1,r06}, the full quasi-hole wave 
functions, which also exhibit the full dependence on the quasi-holes coordinates,
have only been known for up to four quasi-holes in the MR state 
(see \cite{nw1}) and for an arbitrary number
of quasi-holes in the Laughlin states. In this paper, we fill in this gap, 
and calculate the full (four) quasi-hole wavefunctions for the RR 
states and their spin-singlet analogues. In this section, we provide the 
details of the $k=3, M=0$ case of the RR states. We give the 
general $k,M$ results in the appendix \ref{su2app}, where we also
give the details of the fusion rules, OPE's, some general parafermion 
correlators, and the braid relations.

Before we turn to the quasi-hole wavefunctions, we first give the 
wavefunction of the $k=3$ RR states without quasi-holes. In this case
the number of electrons (even though for $M=0$, the particles are bosons,
we will refer to them as electrons) has to be a multiple of 3. It was shown in 
\cite{cgt1} that the following wavefunction is equivalent to the wavefunction
presented in the original paper \cite{rr2}. Divide the electrons into three
groups $S_a$, $a=1,2,3$ of equal size. For each group, we write a
Laughlin factor
\begin{equation}
\Psi^2_{S_a} (\{z\})= \prod_{\substack{i<j\\ i,j \in S_a}} (z_i-z_j)^2 \ .
\end{equation} 
To obtain the RR wavefunction, we sum over all inequivalent ways
to divide the electrons into three groups of equal size
\begin{equation}
\label{rrel}
\Psi_{\rm RR}^{k=3} (\{z\}) = \frac{1}{\cN} \, \sum_{S_1,S_2,S_3} \Bigl[
\Psi^2_{S_1} \Psi^2_{S_2} \Psi^2_{S_3} 
\Bigr] \ ,
\end{equation}
The normalization is $\cN= 3^\frac{N}{2}/3!$, chosen
consistently with the operator product expansion of the parafermion
fields. In effect, the sum amounts to symmetrization of all coordinates.

The $k=3$ clustering property is manifest from the equation (\ref{rrel}):
{}from the wavefunction \eqref{rrel} it is clear that we can put three 
electrons at the same location, without obtaining zero, because there will 
always be a term in the summation for which the three electrons belong to
different groups. Putting four or more electrons at the same location
gives a vanishing wavefunction. 

\subsection{The CFT formulation}

The $k=3$ RR wavefunctions for $N$ electrons and $n$ quasi-holes can 
be expressed in terms of a parafermionic correlator in the following way
\bea
\lefteqn{\Psi_{\rm RR} (w_1,\ldots,w_n;z_1,\ldots,z_N) =}
\nonu &&
\left<
\sigma_1 (w_1) \cdots \sigma_1 (w_n)
\psi_1 (z_1)\cdots\psi_1 (z_N)
\right>^{(0,1)} 
\nonu && 
\times
\prod_{i<j}(z_i-z_j)^{\frac{2}{3}+M} 
\prod_{i,j}(w_i-z_j)^{\frac{1}{3}}
\prod_{i<j}(w_i-w_j)^{\frac{1}{6}-\frac{M}{2(3M+2)}} \ .
\label{k3rrcor}
\eea

{}From the fusion rules, it follows that the number of electrons
$N$ and $n$ have to satisfy the relation $2N+n = 0 \mmod 3$
in order for the correlator to be non-zero. In addition, a state with
only one quasi-hole is impossible.

\subsection{The quasi-hole wavefunctions}
\label{qhwf23}

We focus on the case of four quasi-holes. In this case, there are
two fusion channels (labeled by $(0)$ and $(1)$)
for the parafermion  correlator, namely
$\sigma_1 \sigma_1 \sim \psi_1$ for the $(0)$ channel and
$\sigma_1 \sigma_1 \sim \sigma_2$ for the $(1)$ channel.

Following \cite{cgt1} and specifying to $N=3 r +1$ electrons, with $r$ an 
integer, we can write the following ansatz for the wavefunction for four 
quasi-holes, \eqref{k3rrcor},
\bea
\lefteqn{\Psi^{(0,1)}_{\rm RR} (w_1,w_2,w_3,w_4;z_1,\ldots,z_N) =}
\nonu &&
A^{(0,1)} (\{w\}) \Psi_{12,34} (\{w\}, \{z\}) +
B^{(0,1)} (\{w\}) \Psi_{13,24} (\{w\}, \{z\}) \ . 
\label{meq}
\eea
Throughout the paper, we will choose the phase of wavefunctions in
such a way that the function $A^{(0)} (\{ w\})$ has no phase.

To specify the functions $\Psi_{12,34}$ and $\Psi_{13,24}$, we divide
the electrons in three groups in such a way that $S_1$ contains
$(N-1)/3+1$ electrons and $S_2$ and $S_3$ contain $(N-1)/3$
electrons.
 
Splitting the four quasiholes into two groups, we have
\bea
\label{k3bs}
&& \Psi_{12,34} = \frac{1}{\cN} \sum_{S_1,S_2,S_3} \Bigl[
\prod_{i\in S_2} (z_i-w_1)(z_i-w_2)
\prod_{j\in S_3} (z_j-w_3)(z_j-w_4)
\Psi^2_{S_1} \Psi^2_{S_2} \Psi^2_{S_3} \Bigr] 
\nonu
&& \Psi_{13,24} = \frac{1}{\cN} \sum_{S_1,S_2,S_3} \Bigl[
\prod_{i\in S_2} (z_i-w_1)(z_i-w_3)
\prod_{j\in S_3} (z_j-w_2)(z_j-w_4)
\Psi^2_{S_1} \Psi^2_{S_2} \Psi^2_{S_3} \Bigr] \ ,
\eea
with
$\cN = 3^{\frac{N}{2}}$.
In the following, we will use the case of $N=4$ electrons
to determine the functions $A^{(0,1)}(\{w\})$ and $B^{(0,1)}(\{w\})$.
Note that in the functions \eqref{k3bs}, the quasiholes do not have a zero 
with electrons of the first group before symmetrization. Nevertheless, 
the maximum degree of each $z$ is the same, because the number of electrons 
in the first group is one bigger in comparison to the other two groups.

There is a third possible way of dividing the four quasiholes into 
two groups. However, this does not give an independent function,
because we have the following relation
\begin{equation}
\label{qhrel}
\Psi_{14,23} = x \, \Psi_{12,34} + (1-x) \Psi_{13,24} \ .
\end{equation}
Relations of this kind were first studied by Nayak and Wilczek in \cite{nw1}, 
in order to reduce the overcomplete basis to a linear independent one. 

We should note that an (explicitly) independent basis for the Read-Rezayi 
states (for arbitrary $k$) was recently formulated by Read \cite{r06}
(building on results presented in \cite{aks1}). However,
for our present purposes, namely deriving the full quasi-hole
wavefunctions and studying the braid behaviour of the quasi-holes, it is 
more convenient to use the basis states \eqref{k3bs}, 
because the formulas and the transformation properties under the braiding
of quasi-holes are simpler.

The strategy to obtain the functions $A^{(p)}$ and $B^{(p)}$ in \eqref{meq}
is as follows. Making use of OPE's, including the OPE coefficients,
we reduce the correlator in the wavefunction \eqref{k3rrcor} to 
correlators involving just four $\sigma_{1,2}$ fields. The latter can 
be extracted from the results in \cite{kz1}.
In particular we consider the following limits
\bea
& (I) \qquad & 
z_2 \to z_1, \, z_4 \to z_3, \, z_3 \to w_4, \, z_1 \to w_2,
\nonu
& (II) \qquad & 
z_2 \to z_1, \, z_4 \to z_3, \, z_3 \to w_4, \, z_1 \to w_3 \, . 
\eea
In both limits, we first fuse the two pairs of $\psi_1$ fields to two $\psi_2$ 
fields. Fusing these $\psi_2$ fields with $\sigma_1$ fields, we obtain
$\sigma_2$ fields (see section \ref{afus} for more details about the fusion
rules). Thus, we obtain the following correlators
\begin{align}
\lim_{(I)} & 
\left<\sigma_1 \sigma_1 \sigma_1 \sigma_1
\psi_1 \psi_1 \psi_1 \psi_1\right>^{(0,1)}
\propto  \left<\sigma_1 \sigma_2 \sigma_1\sigma_2 \right>^{(0,1)} 
\nonu
\lim_{(II)} & 
\left<\sigma_1 \sigma_1 \sigma_1 \sigma_1
\psi_1 \psi_1 \psi_1 \psi_1\right>^{(0,1)}
\propto \left<\sigma_1\sigma_1 \sigma_2 \sigma_2 \right>^{(0,1)}
\end{align}

Keeping track of the various OPE coefficients and various
phases is crucial in this procedure. Many of these
coefficients for the $\ZZ_k$ parafermion CFT are given in
\cite{fz1}. The other coefficients we need were obtained from
consistency relations on the four point correlators. We give
these coefficients in appendix \ref{aope}.

With respect to the various phases which need to be taken into
account we note the following. To be able to fuse the fields $\psi_2$
with the appropriate spin field, one has to move the field $\psi_2 (z_1)$
in between the $\sigma_1$ fields. The resulting phase depends on the
fusion channel, and can be obtained from the OPE of $\psi_2$ and the
fusion channel under consideration.
As an example, in the limits above
we need to find the phase associated with
$\sigma_1 (w_3) \sigma_1 (w_4) \psi_2 (z_1) =
(-1)^\alpha \psi_2 (z_1) \sigma_1 (w_3) \sigma_1 (w_4)$. 
In the $(0)$ channel, we
have $\sigma_1 \sigma_1 \sim \psi_1$, 
and the phase $\alpha=-\frac{4}{3}$ follows from
$$\psi_1 (z) \psi_2 (z') \sim (z-z')^{-\frac{4}{3}} 
\sim (-1)^{-\frac{4}{3}} (z'-z)^{-\frac{4}{3}}
\sim (-1)^{-\frac{4}{3}} \psi_2 (z') \psi_1(z) \ ,$$
while in the $(1)$ channel we have $\sigma_1 \sigma_1 \sim \sigma_2$
and the phase $\alpha=-\frac{1}{3}$ follows from 
$$\sigma_2 (z) \psi_2 (z') \sim \vep(z') (z-z')^{-\frac{1}{3}} \sim
(-1)^{-\frac{1}{3}} \vep(z') (z'-z)^{-\frac{1}{3}} \sim (-1)^{-\frac{1}{3}} 
\psi_2 (z') \sigma_2 (z) \ . $$

Taking the phases and OPE coefficients into account, we obtain the limits 
$(I)$ and $(II)$
of the wavefunction $\Psi^{(p)}_{\rm RR}$, with $p=0,1$, eq. \eqref{k3rrcor}
\bea &&
\lim_{(I)} \Psi^{(p)}_{\rm RR} = -(-1)^{p} \frac{4}{9}
(w_{12}w_{34})^{\frac{5}{6}} (1-x)^\frac{1}{6}
\left<\sigma_1 (w_1) \sigma_2 (w_2) \sigma_1 (w_3) \sigma_2 (w_4) \right>^{(p)}
(w_{42}^4 w_{14} w_{32}) 
\nonu &&
\lim_{(II)} \Psi^{(p)}_{\rm RR} = (-1)^{p} \frac{4}{9}
(w_{12}w_{34})^{\frac{5}{6}} x^{-\frac{2}{3}} (1-x)^\frac{5}{6}
\left<\sigma_1 (w_1) \sigma_1 (w_2) \sigma_2 (w_3) \sigma_2 (w_4) \right>^{(p)}
(w_{34}^4 w_{14} w_{32}) \ ,
\nonu &&
\eea
where we used the notation $w_{ij}=(w_i-w_j)$ and the 
following convention for the anharmonic ratio $x$
\begin{align}
x &= \frac{(w_{12})(w_{34})}{(w_{14})(w_{32})} &
1-x &= \frac{(w_{13})(w_{42})}{(w_{14})(w_{32})} &
\frac{x}{1-x} &= \frac{(w_{12})(w_{34})}{(w_{13})(w_{42})} \ , 
\end{align}
which is the same convention as used in \cite{kz1}, but differs from the one
used in \cite{fz1}.
The explicit form of correlators 
$\left<\sigma_1 \sigma_2 \sigma_1 \sigma_2 \right>^{(0,1)}$
and
$\left<\sigma_1 \sigma_1 \sigma_2 \sigma_2 \right>^{(0,1)}$
can be extracted from the results of \cite{kz1}. In 
appendix \ref{scor} we present formulas expressing these correlators
in terms of hypergeometric functions.

Because we can easily take the limits $(I)$ and $(II)$ of the functions
$\Psi_{12,34}$ and $\Psi_{13,24}$ in the case of four electrons,
namely
\begin{align}
\lim_{(I)} \Psi_{12,34} &= -\frac{4}{9} w_{14}w_{32}(w_{42})^4 &
\lim_{(II)} \Psi_{12,34} &= 0 \nonu
\lim_{(I)} \Psi_{13,24} &= 0 &
\lim_{(II)} \Psi_{13,24} &= \frac{4}{9} w_{14}w_{32}(w_{34})^4 \ ,
\end{align}
we finally obtain relations for the functions $A^{(0,1)}$ and $B^{(0,1)}$ 
which are easily solved to give
\begin{align}
A^{(0)} &= (w_{12}w_{34})^{\frac{7}{10}} x^{\frac{3}{10}} 
\cF^{(0)}_1 (x) \nonu
B^{(0)} &= - (w_{12}w_{34})^{\frac{7}{10}} x^{-\frac{7}{10}}(1-x) 
\cF^{(0)}_2 (x) \nonu
A^{(1)} &= -(-1)^{\frac{2}{5}}C (w_{12}w_{34})^{\frac{7}{10}} 
x^{\frac{3}{10}}
\cF^{(1)}_1 (x) \nonu
B^{(1)} &= (-1)^{\frac{2}{5}}C (w_{12}w_{34})^{\frac{7}{10}} 
x^{-\frac{7}{10}} (1-x)
\cF^{(1)}_2 (x) \ .
\end{align}
Here, $C = \frac{1}{2} \sqrt{\frac{\Gamma(\frac{1}{5})\Gamma^3(\frac{3}{5})}
{\Gamma(\frac{4}{5})\Gamma^3 (\frac{2}{5})}}$ and the functions
$\cF^{(p)}_i$, with $p=0,1$ and $i=1,2$ are given in \eqref{fsu23}.

With the explicit form of the functions $A^{(p)}$ and 
$B^{(p)}$, we have completely specified the wavefunction for the $k=3,M=0$ 
Read-Rezayi state in the presence of four quasi-holes. We will give the 
general $k,M$ results in the appendix \ref{su2app}. As an immediate 
application, we can study the behaviour under exchange of quasi-holes, by 
making use of transformation properties of the hypergeometric functions, 
which are also given in appendix \ref{su2app}.

\subsection{Braid behaviour}

To study the braid behaviour under the exchange of quasi-holes, we first 
note that the anharmonic ratio transforms as
$x \mapsto \frac{-x}{1-x}$ for $(1\leftrightarrow 2)$,
$x \mapsto 1-x$ for $(2\leftrightarrow 3)$ and 
$x \mapsto \frac{1}{x}$ for $(1\leftrightarrow 3)$.
In addition, we find that, by making use of \eqref{qhrel},
\begin{align}
\Psi_{12,34} &\xrightarrow{1\leftrightarrow 3} x \, \Psi_{12,34} + (1-
x) \Psi_{13,24} \nonu
\Psi_{13,24} &\xrightarrow{1\leftrightarrow 2} x \, \Psi_{12,34} + (1-x) 
\Psi_{13,24} \ ,
\end{align}
while the other transformations of the $\Psi$'s are clear. 
The braid transformations of the functions $\cF^{(p)}_i$ are
given in appendix \ref{su2br}.
Combining all the results, we find the following braid behaviour
for general $M$
\begin{align}
\Psi^{(p)}_{\rm RR} &\xrightarrow{1\leftrightarrow 2} 
(U_{12})^p_q \Psi^{(q)}_{\rm RR} & 
U_{12} &= (-1)^{-\frac{1}{2}-\frac{M}{2(3M+2)}}
\begin{pmatrix}
(-1)^{-\frac{4}{5}} & 0 \\ 0 & (-1)^{\frac{3}{5}}
\end{pmatrix} \nonu
\Psi^{(p)}_{\rm RR} &\xrightarrow{2\leftrightarrow 3} 
(U_{23})^p_q \Psi^{(q)}_{\rm RR} & 
U_{23} &= (-1)^{-\frac{3}{10}-\frac{3M}{2(3M+2)}}
\begin{pmatrix}
\tau & (-1)^{-\frac{2}{5}}\sqrt{\tau} \\ 
(-1)^{\frac{2}{5}}\sqrt{\tau} & - \tau \end{pmatrix} \nonu
\Psi^{(p)}_{\rm RR} &\xrightarrow{1\leftrightarrow 3} 
(U_{13})^p_q \Psi^{(q)}_{\rm RR} &  
U_{13} &= (-1)^{\frac{3}{10}-\frac{M}{2(3M+2)}} 
\begin{pmatrix} \tau & (-1)^{\frac{1}{5}}\sqrt{\tau} \\ 
-\sqrt{\tau} & (-1)^{\frac{1}{5}} \tau \end{pmatrix}
\label{rrk3braid}
\end{align}
where we use $\tau$ to denote the inverse of the golden ratio, 
i.e. $\tau = \frac{\sqrt{5}-1}{2}$. These matrices are unitary, and 
satisfy $U_{13} = U_{23} \cdot U_{12} \cdot U_{23}^{-1}$. 
Note that if we evenly distribute the phases of the off-diagonal
elements, which amounts to a `gauge' transformation, the matrices
$U_{12}$, $U_{23}$ and $U_{13}$ become proportional to the (inverses of
the) $R$, $F$ and $B$ matrices of the `Fibonacci' anyons, as displayed in
eq.\ (\ref{eq:RFB}).

\section{Quasi-hole wavefunctions for a paired spin-singlet state}
\label{sec:asstates}

\subsection{A paired spin-singlet quantum Hall state}

In the search for a topological quantum liquid suited for universal 
quantum computation, the paired spin-singlet quantum Hall state of
\cite{as1} imposes itself as a natural candidate. This state is the 
$k=2$ member of a series of spin-singlet states introduced and studied
in \cite{as1,arrs}. In many ways, these states are direct extensions of
the Read-Reazayi states to spin-full (spin-1/2) fermions.

The simplest fermionic spin-singlet state (with $M=1$) has filling 
fraction $\nu=4/7$. At this particular filling spin-singlet quantum Hall 
states have been observed \cite{cykcgbw}, but their precise nature has 
never been determined.

The braiding properties of the paired spin-singlet state are essentially
more complicated than those of the paired spin-polarized (Moore-Read) state.
In fact, we will show that the braiding in the paired ($k=2$) spin-singlet
state is similar to that in the $k=3$ Read-Rezayi state, the similarity
being due to what is known as level-rank duality between the affine Lie 
algebras $su(2)_3$ and $su(3)_2$. With this, the $k=2$
state offers the perspective of universal topological quantum computation 
in a paired quantum Hall state. This being enough excitement for us now, 
we shall in this paper not address the $k>2$ spin-singlet states. Their
wavefunctions can be obtained using the methods described in this paper. 

\subsection{The CFT formulation}

The various wavefunctions for the $k=2$ spin-singlet state in the 
presence of quasi-holes are all expressed as correlators in a CFT. For
$M=0$ the CFT is precisely the (chiral) $SU(3)_2$ WZW model while
for $M\neq 0$ we have a deformation thereof, with a modified 
compactification radius of the charge boson $\varphi_c$, see \cite{arrs}.
In all cases, the theory is conveniently represented as a
product of the $su(3)_2$ parafermions, as introduced by Gepner in
\cite{gep1} and the CFT of spin and charge bosons $\varphi_s$ and 
$\varphi_c$.

The fundamental quasi-holes over this quantum Hall state come in 
different types. One type has spin-1/2 and charge $1/(4M+3)$; a second 
option is to have spin-less quasi-holes of charge $2/(4M+3)$.
For $M>0$ the latter have the smaller scaling dimension and are 
thereby the most `relevant' in the sense of scaling arguments.
We expect that experimental protocols for the detection of non-abelian
statistics and for quantum computation will be most easily implemented
using the spin-less quasi-holes.

Using the quantum Hall - CFT connection, we can write the wavefunction for a
state with the number of quasi-holes of the various types specified
as $n_\up$, $n_\down$, $n_3$. These numbers satisfy
\be
N_\up + n_\up = N_\down + n_\down \ ,
\ee
in order for the state to be a spin-singlet and
\be
3 (N_\ua + N_\da) + (n_\ua + n_\da) + 2n_3 = 0 \mmod 4 
\ee
such that the fields can be fused to the identity sector.
The case of only one quasi-hole is an exception, because
a state with only one quasi-hole is impossible.

In full generality the quasi-hole wavefunction reads \cite{arrs} 
\begin{eqnarray}
\lefteqn{  
\Psi_{\rm AS}^{M}
(w_1^\uparrow,\ldots,w_{n_\uparrow}^\uparrow;
 w_1^\downarrow,\ldots,w_{n_\downarrow}^\downarrow;
 w_1,\ldots,w_{n_3};
 z_1^\uparrow,\ldots,z_{N_\uparrow}^\uparrow;
z_1^\downarrow,\ldots,z_{N_\downarrow}^\downarrow) = } 
\nonu &&
\hspace{-9pt}
\langle 
\sigma_\uparrow(w_1^\uparrow) \ldots 
\sigma_\uparrow(w_{n_\uparrow}^\uparrow) 
\sigma_\downarrow(w_1^\downarrow) \ldots 
\sigma_\downarrow(w_{n_\downarrow}^\downarrow)
\sigma_3(w_1) \ldots 
\sigma_3(w_{n_3})
\psi_1(z_1^\uparrow) \ldots \psi_1(z_{N_\uparrow}^\uparrow)
\psi_2(z_1^\downarrow) \ldots \psi_2(z_{N_\downarrow}^\downarrow)
\rangle 
\nonu && 
\hspace{-9pt}
\times 
\left[ \wtPsi_{\rm H}^{(2,2,1)}(z_i^\uparrow;z_j^\downarrow) \right]^{1/2} \,
\wtPsi^M_{\rm L}(z_i^\uparrow;z_j^\downarrow) 
\prod_{i,j} (z^\uparrow_i-w^\uparrow_j)^{\frac{1}{2}} 
\prod_{i,j} (z^\downarrow_i-w^\downarrow_j)^{\frac{1}{2}} 
\prod_{i,j} (z^\uparrow_i-w_j)^{\frac{1}{2}} 
\prod_{i,j} (z^\downarrow_i-w_j)^{\frac{1}{2}} 
\nonu && 
\hspace{-9pt}
\times
\prod_{i<j}(w^\uparrow_i-w^\uparrow_j)^{\frac{1}{3}-\frac{M}{3(4M+3)}}
\prod_{i,j}(w^\uparrow_i-w^\downarrow_j)^{-\frac{1}{6}-\frac{M}{3(4M+3)}}
\prod_{i<j}(w^\downarrow_i-w^\downarrow_j)^{\frac{1}{3}-\frac{M}{3(4M+3)}}
\nonu &&
\hspace{-9pt}
\times
\prod_{i,j}(w^\uparrow_i-w_j)^{\frac{1}{6}-\frac{2M}{3(4M+3)}}
\prod_{i,j}(w^\downarrow_i-w_j)^{\frac{1}{6}-\frac{2M}{3(4M+3)}}
\prod_{i<j}(w_i-w_j)^{\frac{1}{3}-\frac{4M}{3(4M+3)}}
\ .
\label{eq:psiqh}
\end{eqnarray}
Below we present explicit formulas for these wavefunctions
in the special cases of four quasi-holes with (i) $n_3=4$,
(ii) $n_\uparrow = n_\downarrow=2$, (iii) $n_\uparrow=4$
and (iv) $n_\ua=3 \ , n_\da =1$.
To avoid clutter, we write the expressions for $M=0$; the
braid relations will be specified for general $M$.

\subsection{Evaluating quasi-hole wavefunctions}
\label{sec:qhsu32}

We first give the wavefunction without any quasi-holes in the
form given in \cite{sal1}, see also \cite{cgt1}. Assuming 
$N_\uparrow=N_\downarrow$ both even we have
(the sum is over all independent ways of dividing the electrons in two groups, both
containing $N_\ua/2$ spin-up electrons and $N_\da/2$ spin-down
electrons; in effect this amounts to symmetrization over the spin-up
and spin-down electrons)
\be
\label{asel}
\Psi_{\rm AS}^{M=0} = \frac{1}{\cN}
\sum_{S_1,S_2} \Psi^{221}_{S_1}(\ziu,\zjd) \Psi^{221}_{S_2}(\zku,\zld) \ ,
\ee
with
$\cN = 2^{\frac{N_\ua+N_\da}{2}}/2$
and with $\Psi^{221}_{S_a}$ denoting the 221 state restricted 
to $\ziu,\zjd\in S_a$
\be
\Psi^{221}_{S_a} (\ziu,\zjd) =
\prod_{\substack{i<j\\ i,j\in S_a}} (\ziu-\zju)^2
\prod_{\substack{i'<j'\\ i',j' \in S_a}} (\zid-\zjd)^2
\prod_{\substack{i,j'\\ i,j' \in S_a}} (\ziu-\zjd) \ .
\ee
Note that from now on, we will put a prime on the index of spin-down
particles. In some cases we will drop the arrows on the quasi-hole
coordinates.

The validity of expression \eqref{asel} is most easily
understood from the characterization of the paired spin-singlet as the
maximal-degree zero-energy eigenfunction of a specific 3-body hamiltonian.

\subsubsection{The case $n_3=4$}

The wavefunction takes the form
\begin{eqnarray}
\lefteqn{  
\Psi_{\rm AS}[3333]
( w_1,w_2,w_3,w_4;
  z_1^\uparrow,\ldots,z_{N_\uparrow}^\uparrow;
  z_{1'}^\downarrow,\ldots,z_{N'_\downarrow}^\downarrow) = } 
\nonu &&
\langle 
\sigma_3(w_1) \sigma_3(w_2) \sigma_3(w_3) \sigma_3(w_{4})
\psi_1(z_1^\uparrow) \ldots \psi_1(z_{N_\uparrow}^\uparrow)
\psi_2(z_{1'}^\downarrow) \ldots \psi_2(z_{N'_\downarrow}^\downarrow)
\rangle 
\nonu && 
\times 
\left[ \wtPsi_{\rm H}^{(2,2,1)}(z_i^\uparrow;z_{j'}^\downarrow) \right]^{1/2} 
\, \prod_{i,j} (z^\uparrow_i-w_j)^{\frac{1}{2}} 
\prod_{i',j} (z^\downarrow_{i'}-w_j)^{\frac{1}{2}} 
\prod_{i<j}(w_i-w_j)^{\frac{1}{3}} \ .
\label{eq:psiqh3333}
\end{eqnarray}
Defining
\bea
\lefteqn{
\Psi_{ab,cd}= \frac{1}{2 {\cal N}}
\sum_{S_1,S_2}
\left[ \prod_{i,j'\in S_1} (\ziu - w_a)(\zjd - w_a)
                           (\ziu - w_b)(\zjd - w_b) 
\right]
\Psi^{221}_{S_1}(\ziu,\zjd)}
\nonu && 
\qquad \qquad \quad \times
\left[ \prod_{k,l'\in S_2} (\ziu - w_c)(\zjd - w_c)
                           (\ziu - w_d)(\zjd - w_d) 
\right] 
\Psi^{221}_{S_2}(\zku,\zld) \ .
\eea
we propose the following expression
\bea
\lefteqn{\Psi_{\rm AS}^{(0,1)}[3333]
(w_1,w_2,w_3,w_4;z^\ua_1,\ldots,z^\ua_{N_\ua};z^\da_{1'},\ldots,z^\da_{N'_\da})=}
\nonu &&
A^{(0,1)}[3333](\{w\}) \Psi_{12,34} (\{w\},\{z\})
+
B^{(0,1)}[3333](\{w\}) \Psi_{13,24} (\{w\},\{z\}) \ .
\label{eq:master3333}
\eea
Following steps that are similar to those presented in
section \ref{qhwf23}, we can determine the coefficients
in the master formula eq. (\ref{eq:master3333}). The
particular limits we employ are
\bea
& (I) \qquad & z_1^\uparrow \to z_2^\uparrow, \, 
               z_{1'}^\downarrow \to z_{2'}^\downarrow, \, 
\nonu
& (II) \qquad & z_{1'}^\downarrow \to z_{2'}^\downarrow, \, 
                z_1^\uparrow \to w_1, z_2^\uparrow \to w_2  \, . 
\eea
They give
\begin{align}
\lim_{(I)} & 
\left<\sigma_3 \sigma_3 \sigma_3 \sigma_3
\psi_1 \psi_1 \psi_2 \psi_2\right>^{(0,1)}
\propto  \left<\sigma_3 \sigma_3 \sigma_3 \sigma_3 \right>^{(0,1)} 
\nonu
\lim_{(II)} & 
\left<\sigma_3 \sigma_3 \sigma_3 \sigma_3
\psi_1 \psi_1 \psi_2 \psi_2\right>^{(0,1)}
\propto \left<\sigma_\ua \sigma_\ua \sigma_3 \sigma_3 \right>^{(0,1)}
\end{align}
leading to the result
\begin{align}
A^{(0)}[3333] &= (w_{12}w_{34})^{\frac{4}{5}} x^{-\frac{2}{15}} 
(1-x)^{\frac{2}{3}} \cF^{(0)}_2 (x) 
\nonu
B^{(0)}[3333] &= (w_{12}w_{34})^{\frac{4}{5}} x^{-\frac{2}{15}}
(1-x)^{\frac{2}{3}} \cF^{(0)}_1 (x) 
\nonu
A^{(1)}[3333] &= - (-1)^{\frac{3}{5}}C (w_{12}w_{34})^{\frac{4}{5}} 
x^{-\frac{2}{15}} (1-x)^{\frac{2}{3}} \cF^{(1)}_2 (x) 
\nonu
\label{eq:four3w}
B^{(1)}[3333] &= - (-1)^{\frac{3}{5}}C (w_{12}w_{34})^{\frac{4}{5}} 
x^{-\frac{2}{15}} (1-x)^{\frac{2}{3}} \cF^{(1)}_1 (x) \ .
\end{align}
Here, $C = \frac{1}{3} \sqrt{\frac{\Gamma(\frac{4}{5})\Gamma^3(\frac{2}{5})}
{\Gamma(\frac{1}{5})\Gamma^3 (\frac{3}{5})}}$ and the functions
$\cF^{(p)}_i(x)$, with $p=0,1$ and $i=1,2$ are given in \eqref{fsu32}.
Note that, while we use the same notation as in section \ref{qhwf23}, 
the actual functions $\cF^{(p)}_i(x)$ and the value of $C$ differ between 
the two cases.

\subsubsection{The case $n_\uparrow=n_\downarrow=2$}

The wavefunction reads
\bea
\lefteqn{  
\Psi_{\rm AS}[\uparrow\uparrow\downarrow\downarrow]
(w_1^\uparrow,w_{2}^\uparrow;
 w_{3'}^\downarrow,w_{4'}^\downarrow;
 z_1^\uparrow,\ldots,z_{N_\uparrow}^\uparrow;
 z_1^\downarrow,\ldots,z_{N_\downarrow}^\downarrow) = } 
\nonu &&
\langle 
\sigma_\uparrow(w_1^\uparrow) \sigma_\uparrow(w_{2}^\uparrow) 
\sigma_\downarrow(w_{3'}^\downarrow) \sigma_\downarrow(w_{4'}^\downarrow)
\psi_1(z_1^\uparrow) \ldots \psi_1(z_{N_\uparrow}^\uparrow)
\psi_2(z_1^\downarrow) \ldots \psi_2(z_{N_\downarrow}^\downarrow)
\rangle 
\nonu && 
\times 
\left[ \wtPsi_{\rm H}^{(2,2,1)}(z_i^\uparrow;z_{j'}^\downarrow) \right]^{1/2}  
\prod_{i,j} (z^\uparrow_i-w^\uparrow_j)^{\frac{1}{2}} 
\prod_{i',j'} (z^\downarrow_{i'}-w^\downarrow_{j'})^{\frac{1}{2}} 
\nonu && 
\times
\prod_{i<j}(w^\uparrow_i-w^\uparrow_j)^{\frac{1}{3}}
\prod_{i,j'}(w^\uparrow_i-w^\downarrow_{j'})^{-\frac{1}{6}}
\prod_{i'<j'}(w^\downarrow_{i'}-w^\downarrow_{j'})^{\frac{1}{3}} \ . 
\label{eq:psiqhupupdownndown}
\eea
The master formula now reads (from now on, we will label the quasi-holes
consecutively, to avoid confusion when braiding quasi-holes; as a reminder, we
will put primes on the labels of the spin-down quasi-holes)
\bea
\lefteqn{
\Psi_{\rm AS}^{(0,1)}[\up\up\down\down]
(w^\ua_1,w^\ua_2;w^\da_{3'},w^\da_{4'};z^\ua_1,\ldots,
z^\ua_{N_\ua};z^\da_{1'},\ldots,z^\da_{N'_\da})=} 
\nonu &&
A^{(0,1)}[\up\up\down\down](\{w\}) \Psi_{13',24'} (\{w\},\{z\})
+
B^{(0,1)}[\up\up\down\down](\{w\}) \Psi_{14',23'} (\{w\},\{z\}) \ .
\label{eq:masterupupdowndown}
\eea
with 
\bea
\lefteqn{\Psi_{13',24'}=
\frac{1}{2 {\cal N}}
\sum_{S_1,S_2}
\left[ \prod_{i,j'\in S_1} (\ziu - w^\ua_1)(\zjd -w^\da_{3'} ) \right]
\Psi^{221}_{S_1}(\ziu,\zjd)}
\nonu && 
\qquad \qquad \quad \times
\left[ \prod_{k,l'\in S_2} (\zku - w^\ua_2)(\zld - w^\da_{4'}) \right]
\Psi^{221}_{S_2}(\zku,\zld) \ .
\eea
and similarly $\Psi_{14',23'}$. Note that in this case there is no 
natural third way to distribute the quasi-holes over $S_1$, $S_2$.

To determine $A^{(0,1)}[\up\up\down\down]$ and 
$B^{(0,1)}[\up\up\down\down]$ we put 
$N_\uparrow=N_\downarrow=2$ and consider the limits
\bea
& (I) \qquad & \zoneu \to \ztwou, \, \zoned \to \ztwod,
\nonu
& (II)\qquad &
\zoneu \to \woneu, \,
\ztwou \to \wtwou, \,
\zoned \to w^\da_{3'}, \,
\ztwod \to w^\da_{4'} \ .
\eea
They give
\begin{align}
\lim_{(I)} & 
\left<\sigma_\uparrow \sigma_\uparrow \sigma_\downarrow \sigma_\downarrow
\psi_1 \psi_1 \psi_2 \psi_2\right>^{(0,1)}
\propto  \left<
\sigma_\uparrow \sigma_\uparrow \sigma_\downarrow \sigma_\downarrow
\right>^{(0,1)} \nonu
\lim_{(II)} & 
\left<\sigma_\uparrow \sigma_\uparrow \sigma_\downarrow \sigma_\downarrow
\psi_1 \psi_1 \psi_2 \psi_2\right>^{(0,1)}
\propto \left<\sigma_3 \sigma_3 \sigma_3 \sigma_3 \right>^{(0,1)} \ .
\end{align}
In the $(0)$ channel this gives the equations
\bea
\lefteqn{(I) \quad\; (A+B)^{(0)}[\up\up\down\down] =
(w_{12}w_{3'4'})^{-\frac{1}{5}} x^{\frac{13}{15}}(1-x)^{-\frac{1}{3}} 
{\cal F}_1^{(0)}(x)}
\nonu
\lefteqn{(II) \quad 
(\sqrt{1-x} \, A + \frac{1}{\sqrt{1-x}} B)^{(0)}[\up\up\down\down] =}
\nonu 
&& \qquad\quad
(w_{12}w_{3'4'})^{-\frac{1}{5}} x^{\frac{13}{15}}(1-x)^{\frac{1}{6}} 
[ {\cal F}_1^{(0)}(x) + {\cal F}_2^{(0)}(x)] \ .
\eea
The equations  in channel $(1)$ have a similar structure. The solutions are
\begin{align}
A^{(0)}[\up\up\down\down] 
&= (w_{12}w_{3'4'})^{-\frac{1}{5}} x^{\frac{13}{15}} 
(1-x)^{-\frac{1}{3}} 
[{\cal F}_1^{(0)}(x) - \frac{1-x}{x} {\cal F}_2^{(0)}(x)]
\nonu
B^{(0)}[\up\up\down\down] 
&= (w_{12} w_{3'4'})^{-\frac{1}{5}} x^{\frac{13}{15}}
(1-x)^{-\frac{1}{3}} 
[ \frac{1-x}{x} {\cal F}_2^{(0)}(x) ]
\nonu
A^{(1)}[\up\up\down\down] 
&= -(-1)^{\frac{3}{5}} C
(w_{12}w_{3'4'})^{-\frac{1}{5}} x^{\frac{13}{15}} 
(1-x)^{-\frac{1}{3}} [ {\cal F}_1^{(1)}(x) 
- \frac{1-x}{x} {\cal F}_2^{(1)}(x)]
\nonu
B^{(1)}[\up\up\down\down] 
&= -(-1)^{\frac{3}{5}} C
(w_{12}w_{3'4'})^{-\frac{1}{5}} x^{\frac{13}{15}}
(1-x)^{-\frac{1}{3}} 
[ \frac{1-x}{x} {\cal F}_2^{(1)}(x)] \, .
\end{align}

Interchanging the positions of $\sigma_\up(w_2)$ and
$\sigma_\down(w_{3'})$ gives a different basis for the
four quasi-hole wavefunctions in this sector. With
\bea
\lefteqn{\Psi_{\rm AS}^{(0,1)}[\up\down\up\down]
(w^\ua_1,w^\da_{2'};w^\ua_3,w^\da_{4'};z^\ua_1,\ldots,
z^\ua_{N_\ua};z^\da_{1'},\ldots,z^\da_{N'_\da})=} 
\nonu &&
A^{(0,1)}[\up\down\up\down](\{w\}) \Psi_{12',34'} (\{w\},\{z\})
+
B^{(0,1)}[\up\down\up\down](\{w\}) \Psi_{14',32'} (\{w\},\{z\}) \ .
\label{eq:masterupdownupdown}
\eea
the  coefficients are found to be
\begin{align}
A^{(0)}[\up\down\up\down]
&= (-1)^{\frac{1}{2}}
(w_{12'}w_{34'})^{-\frac{1}{5}} x^{-\frac{2}{15}} 
(1-x)^{\frac{2}{3}} 
[{\cal F}_2^{(0)}(x) -\frac{x}{1-x} {\cal F}_1^{(0)}(x)]
\nonu
B^{(0)}[\up\down\up\down] 
&= (-1)^{\frac{1}{2}}
(w_{12'}w_{34'})^{-\frac{1}{5}} x^{-\frac{2}{15}}
(1-x)^{\frac{2}{3}} 
[ \frac{x}{1-x} {\cal F}_1^{(0)}(x) ]
\nonu
A^{(1)}[\up\down\up\down] 
&= (-1)^{\frac{1}{10}} C
(w_{12'}w_{34'})^{-\frac{1}{5}} x^{-\frac{2}{15}} 
(1-x)^{\frac{2}{3}} [ {\cal F}_2^{(1)}(x) 
- \frac{x}{1-x} {\cal F}_1^{(1)}(x)]
\nonu
B^{(1)}[\up\down\up\down] 
&= (-1)^{\frac{1}{10}} C
(w_{12'}w_{34'})^{-\frac{1}{5}} x^{-\frac{2}{15}}
(1-x)^{\frac{2}{3}} 
[ \frac{x}{1-x} {\cal F}_1^{(1)}(x)] \, .
\end{align}

\subsubsection{The case $n_\uparrow=4$}

The wavefunction takes the form
\begin{eqnarray}
\lefteqn{  
\Psi_{\rm AS}[\ua\ua\ua\ua]
(w^\ua_1,w^\ua_2,w^\ua_3,w^\ua_{4};
  z_1^\uparrow,\ldots,z_{N_\uparrow}^\uparrow;
  z_{1'}^\downarrow,\ldots,z_{N'_\downarrow}^\downarrow) = } 
\nonu &&
\langle 
\sigma_\ua(w^\ua_1) \sigma_\ua(w^\ua_2) \sigma_\ua(w^\ua_3) \sigma_\ua(w^\ua_{4})
\psi_1(z_1^\uparrow) \ldots \psi_1(z_{N_\uparrow}^\uparrow)
\psi_2(z_{1'}^\downarrow) \ldots \psi_2(z_{N'_\downarrow}^\downarrow)
\rangle 
\nonu && 
\times 
\left[ \wtPsi_{\rm H}^{(2,2,1)}(z_i^\uparrow;z_{j'}^\downarrow) \right]^{1/2} \, 
\prod_{i,j} (z^\uparrow_i-w^\ua_j)^{\frac{1}{2}} 
\prod_{i<j}(w^\ua_i-w^\ua_j)^{\frac{1}{3}} \ .
\label{eq:psiqhupupupup}
\end{eqnarray}
{}From now on, we will drop the up-arrow on the quasi-hole coordinates
$\{w_i\}$. Defining
\bea
\lefteqn{
\Psi_{ab,cd}= \frac{1}{2 \cN}
\sum_{S_1,S_2}
\left[ \prod_{i\in S_1} (\ziu - w_a)
                           (\ziu - w_b) 
\right]
\Psi^{221}_{S_1}(\ziu,\zjd)}
\nonu && \qquad \qquad \quad \times
\left[ \prod_{i\in S_2} (\ziu - w_c)
                           (\ziu - w_d)
\right] 
\Psi^{221}_{S_2}(\zku,\zld) \ ,
\eea
we propose the following expression
\bea
\lefteqn{\Psi_{AS}^{(0,1)}[\ua\ua\ua\ua] (w_1,w_2,w_3,w_4;
z_1^\uparrow,\ldots,z_{N_\uparrow}^\uparrow;
z_{1'}^\downarrow,\ldots,z_{N'_\downarrow}^\downarrow)
=}
\nonu &&
A^{(0,1)}[\ua\ua\ua\ua](\{w\}) \Psi_{12,34} (\{w\},\{z\})
+
B^{(0,1)}[\ua\ua\ua\ua](\{w\}) \Psi_{13,24} (\{w\},\{z\}) \ .
\label{eq:masterupupupup}
\eea
To determine the coefficients in this master formula,
we set $N_\ua = 2$, $N_\da=6$ and take the two
limits 
\bea
& (I) \qquad & z^\da_{1'} \to z^\ua_1, \, z^\da_{2'} \to z^\ua_2, \,
z^\da_{4'} \to z^\da_{3'}, \, z^\da_{6'} \to z^\ua_{5'}, \,
z^\ua_1 \to w_2, \, z^\ua_2 \to w_4  
\nonu   
& (II) \qquad & z^\da_{1'} \to z^\ua_1, \, z^\da_{2'} \to z^\ua_2, \,
z^\da_{4'} \to z^\da_{3'}, \, z^\da_{6'} \to z^\ua_{5'}, \,
z^\ua_1 \to w_3, \, z^\ua_2 \to w_4 \ ,
\eea
which give
\begin{align}
\lim_{(I)} &
\langle \sigma_\ua \sigma_\ua \sigma_\ua \sigma_\ua 
\psi_1 \psi_1 \psi_2 \psi_2 \psi_2 \psi_2 \psi_2 \psi_2 \rangle^{(0,1)} 
\propto \langle \sigma_\ua \sigma_\da \sigma_\ua \sigma_\da \rangle^{(0,1)} \nonu
\lim_{(II)} &
\langle \sigma_\ua \sigma_\ua \sigma_\ua \sigma_\ua 
\psi_1 \psi_1 \psi_2 \psi_2 \psi_2 \psi_2 \psi_2 \psi_2 \rangle^{(0,1)}
\propto \langle \sigma_\ua \sigma_\ua \sigma_\da \sigma_\da \rangle^{(0,1)} .
\end{align} 
Using the spin field correlators given in eq.\ (\ref{eq:fourspin}) we find
\begin{align}
\label{fourupw}
A^{(0)}[\ua\ua\ua\ua] &=
(w_{12}w_{34})^{\frac{4}{5}} x^{-\frac{2}{15}} (1-x)^{\frac{2}{3}} \cF^{(0)}_2 (x) \nonu
B^{(0)}[\ua\ua\ua\ua] &=
(w_{12}w_{34})^{\frac{4}{5}} x^{-\frac{2}{15}} (1-x)^{\frac{2}{3}} \cF^{(0)}_1 (x) \nonu
A^{(1)}[\ua\ua\ua\ua] &= -(-1)^{\frac{3}{5}} C (w_{12}w_{34})^{\frac{4}{5}}
x^{-\frac{2}{15}} (1-x)^{\frac{2}{3}} \cF^{(1)}_2 (x) \nonu
B^{(1)}[\ua\ua\ua\ua] &= -(-1)^{\frac{3}{5}} C (w_{12}w_{34})^{\frac{4}{5}} 
x^{-\frac{2}{15}} (1-x)^{\frac{2}{3}} \cF^{(1)}_1 (x) \ .
\end{align}

Because the spin-up quasi-holes have to satisfy exactly the same braid 
properties as the spin-less quasi-holes for $M=0$, it is not surprising at 
all that the functional form of the functions \eqref{fourupw} is exactly the 
same as in \eqref{eq:four3w}. 

\subsubsection{The case $n_\uparrow=3$, $n_\da=1$}

The wavefunction is
\begin{eqnarray}
\lefteqn{  
\Psi_{\rm AS}[\ua\ua\ua\da]
( w^\ua_1,w^\ua_2,w^\ua_3;w^\da_{4'};
  z_1^\uparrow,\ldots,z_{N_\uparrow}^\uparrow;
  z_{1'}^\downarrow,\ldots,z_{N'_\downarrow}^\downarrow) = } 
\nonu &&
\langle 
\sigma_\ua(w^\ua_1) \sigma_\ua(w^\ua_2)
\sigma_\ua(w^\ua_3) \sigma_\ua(w^\da_{4'})
\psi_1(z_1^\uparrow) \ldots \psi_1(z_{N_\uparrow}^\uparrow)
\psi_2(z_{1'}^\downarrow) \ldots \psi_2(z_{N'_\downarrow}^\downarrow)
\rangle 
\nonu && 
\times 
\left[ \wtPsi_{\rm H}^{(2,2,1)}(z_i^\uparrow;z_{j'}^\downarrow) \right]^{1/2} 
\, \prod_{i,j} (z^\uparrow_i-w^\ua_j)^{\frac{1}{2}}
\prod_{i'} (z^\downarrow_{i'}-w^\da_{4'})^{\frac{1}{2}} \ .
\nonu && 
\times 
\prod_{i<j}(w^\ua_i-w^\ua_j)^{\frac{1}{3}}
\prod_{i}(w^\ua_i-w^\da_{4'})^{-\frac{1}{6}} \ .
\label{eq:psiqhupupupdown}
\end{eqnarray}
In this case, we will in general have $N_\ua = 2 r +1$ spin-up electrons
and $N_\da=2 r +3$ spin-down electrons, with $r$ an integer. 
Thus, to define the functions
$\Psi_{ab,cd}$, we will divide the electrons in two groups, where the first
group $S_1$ contains $(N_\ua-1)/2$ spin-up electrons and $(N_\da-1)/2+1$
spin-down electrons. The second group $S_2$ has the remaining
$(N_\ua-1)/2+1$ spin-up electrons, and the remaining $(N_\da-1)/2$
spin-down electrons. We can now define
\bea
\lefteqn{
\Psi_{ab,cd'}= \frac{1}{\cN'}
\sum_{S_1,S_2}
\left[ \prod_{i\in S_1} (\ziu - w^\ua_a) (\ziu - w^\ua_b) 
\right]
\Psi^{221}_{S_1}(\ziu,\zjd)}
\nonu && \qquad \qquad \quad \times
\left[ \prod_{i,j'\in S_2} (\ziu - w^\ua_c)(\zjd - w^\da_d)
\right] 
\Psi^{221}_{S_2}(\zku,\zld) \ ,
\eea
with the normalization
$\cN' = 2^{\frac{N_\ua+1}{2}}2^{\frac{N_\da+1}{2}}$, where the sum is 
over all ways 
of dividing the electrons into the two groups. Note that we again 
have three different ways of splitting the quasi-holes. However, 
the relation between them differs form the `usual' relation, because 
we now have
\be
\label{altpsi}
\Psi_{23,14'} =  x \frac{w_{14'}}{w_{34'}}\Psi_{12,34'} -
(1-x) \frac{w_{14'}}{w_{4'2}} \Psi_{13,24'} \ .
\ee

The master formula now reads
\bea
\lefteqn{\Psi_{AS}^{(0,1)}[\ua\ua\ua\da] (w^\ua_1,w^\ua_2,w^\ua_3;w^\da_{4'};
z_1^\uparrow,\ldots,z_{N_\uparrow}^\uparrow;
z_{1'}^\downarrow,\ldots,z_{N'_\downarrow}^\downarrow)=}
\nonu &&
A^{(0,1)}[\ua\ua\ua\da](\{w\}) \Psi_{12,34'} (\{w\},\{z\})
+ 
B^{(0,1)}[\ua\ua\ua\da](\{w\}) \Psi_{13,24'} (\{w\},\{z\}) \ .
\label{eq:masterupupupdown}
\eea
Specifying $N_\ua=1$ and $N_\da=3$, and taking the limits
\bea
& (I) \qquad & z^\da_{1'} \to z^\ua_1, \, z^\da_{3'} \to z^\da_{2'}, \,
z^\ua_1 \to w^\ua_2 \nonu   
& (II) \qquad & z^\da_{1'} \to z^\ua_1, \, z^\da_{3'} \to z^\da_{2'}, \,
z^\ua_1 \to w^\ua_3 \ ,
\eea
we obtain the following result
\begin{align}
\label{threeupdownw}
A^{(0)}[\ua\ua\ua\da] &=
\frac{(w_{12}w_{34'})^{\frac{4}{5}}}{w_{34'}} x^{-\frac{2}{15}} 
(1-x)^{\frac{2}{3}} \cF^{(0)}_2 (x) 
\nonu
B^{(0)}[\ua\ua\ua\da] &= -\frac{(w_{12}w_{34'})^{\frac{4}{5}}}{w_{4'2}} 
x^{-\frac{2}{15}} (1-x)^{\frac{2}{3}} \cF^{(0)}_1 (x) 
\nonu
A^{(1)}[\ua\ua\ua\da] &= -(-1)^{\frac{3}{5}} C 
\frac{(w_{12}w_{34'})^{\frac{4}{5}}}{w_{34'}} x^{-\frac{2}{15}} 
(1-x)^{\frac{2}{3}} \cF^{(1)}_2 (x) 
\nonu
B^{(1)}[\ua\ua\ua\da] &= (-1)^{\frac{3}{5}} C 
\frac{(w_{12}w_{34'})^{\frac{4}{5}}}{w_{4'2}} x^{-\frac{2}{15}} 
(1-x)^{\frac{2}{3}} \cF^{(1)}_1 (x) \ .
\end{align}

Note that in calculating the braid relations, the `extra' factors of 
$w_{34'}$ and $w_{4'2}$ are precisely `compensated' by the additional 
factors in the relation between the $\Psi_{ab,cd'}$'s in eq. \eqref{altpsi}.

\subsection{Braiding relations}

The braid properties of the various four quasi-hole wavefunctions
are easily evaluated using the transformation properties
of the functions $\cF^{(p)}_i(x)$ as specified in section 
\ref{su32braid}. We display them here for general $M$.

For braiding the neutral quasi-holes of charge $2/(4M+3)$, denoted
by the label `3', we obtain the following results
\bea
\label{eq:braid3333}
\Psi_{\rm AS}^{(p)}[3333]
         \stackrel{1 \leftrightarrow 2}{\rightarrow}
         (U_{12})^p_q \Psi_{\rm AS}^{(q)}[3333]
& \quad &
U_{12} = (-1)^{-\frac{2}{3}-\frac{4M}{3(4M+3)}}\left( \begin{array}{cc}
              (-1)^{\frac{4}{5}} & 0 \\
              0 & (-1)^{-\frac{3}{5}} 
                           \end{array} \right)
\\[2mm]
\Psi_{\rm AS}^{(p)}[3333]
         \stackrel{2 \leftrightarrow 3}{\rightarrow}
         (U_{23})^p_q \Psi_{\rm AS}^{(q)}[3333]
& \quad & 
U_{23} = (-1)^{\frac{4}{5}-\frac{4M}{4M+3}}
             \left( \begin{array}{cc}
             \tau & (-1)^{-\frac{3}{5}} \sqrt{\tau} \\
             (-1)^{\frac{3}{5}} \sqrt{\tau} & - \tau 
              \end{array} \right)
\nonu
\Psi_{\rm AS}^{(p)}[3333]
         \stackrel{1 \leftrightarrow 3}{\rightarrow}
         (U_{13})^p_q \Psi_{\rm AS}^{(q)}[3333]
& \quad & 
U_{13} = (-1)^{\frac{8}{15}-\frac{4M}{3(4M+3)}}
             \left( \begin{array}{cc}
             \tau & -(-1)^{-\frac{1}{5}} \sqrt{\tau} \\
             \sqrt{\tau} & (-1)^{-\frac{1}{5}} \tau 
              \end{array} \right) \ . 
\nonumber
\eea
For braiding spin-full quasi-holes, of charge $1/(4M+3)$,
the corresponding matrices are 
\begin{align}
U_{12} &= (-1)^{-\frac{2}{3}-\frac{M}{3(4M+3)}}
\begin{pmatrix}
(-1)^{\frac{4}{5}} & 0 \\ 0 & (-1)^{-\frac{3}{5}}
\end{pmatrix} 
\nonu
U_{23} &= (-1)^{\frac{4}{5}-\frac{M}{4M+3}}
\begin{pmatrix}
\tau & (-1)^{-\frac{3}{5}} \sqrt{\tau} \\ (-1)^{\frac{3}{5}}\sqrt{\tau} & -\tau
\end{pmatrix} 
\nonu
U_{13} &= (-1)^{\frac{8}{15}-\frac{M}{3(4M+3)}}
\begin{pmatrix}
\tau & -(-1)^{-\frac{1}{5}} \sqrt{\tau} \\ \sqrt{\tau} & (-1)^{-\frac{1}{5}}\tau
\end{pmatrix} \ .
\label{eq:braiduuud}
\end{align}
These matrices are found by explicit inspection of the 
wavefunctions for the cases $(n_\ua=3\, , \, n_\da=1)$ and 
$n_\ua=4$. For $(n_\ua=2\, , \, n_\da=2)$ equivalent matrices
are found for situations where the braiding does not mix the
spin-labels, such as $1 \leftrightarrow 2$ for 
$\Psi_{\rm AS}^{(p)}[\up\up\down\down]$ and 
$1 \leftrightarrow 3$ for 
$\Psi_{\rm AS}^{(p)}[\up\down\up\down]$.

The different $M$-dependence between eqs.\ eq.\ (\ref{eq:braid3333}) 
and (\ref{eq:braiduuud}) reflects the fact that for $M\neq 0$ the 
conformal dimension of
$\sigma_3$ differs from that of $\sigma_\up$ and $\sigma_\down$.
For $M=0$ the three quasi-hole types are connected by an $su(3)$
symmetry and the braiding properties are necessarily the same.

By a simple change of basis (`gauge transformation') the matrices
$U_{12}$, $U_{23}$ and $U_{13}$ acquire a form which can be
identified with that of the $R$, $F$ and $B$ matrices for the Fibonacci 
anyons, eq.\ (\ref{eq:RFB}), up to overall $M$-dependent phase factors. 
This establishes the suitability of this particular quantum register 
for universal topological quantum computation.

\section{Quantum group approach}
\label{sec:qg}

In this section, we will use the CFT-quantum group connection
to calculate the braid properties of the quasi-holes, and confirm that
the results from this approach is indeed consistent with the results
obtained here from the explicit wavefunctions for the quasi-hole
states.

The approach used here follows the lines of Slingerland and Bais,
\cite{sb1}, to which we refer for details and more references. More
details on the CFT-quantum group connection can be found in
\cite{gras,fuchs}. For the $su(2)_k$ case in particular, see
\cite{aggs}. More details about the quantum groups themselves
can be found in, for instance, \cite{majid}.

At a basic level, the connection between conformal field theory and
quantum groups states that the braid properties of fields in the
conformal field theory are the same as the braid properties of
particles carrying a quantum group representation. Because the
latter are specified by the $R$-matrix of the quantum group, one
can calculate the braid properties of the quasi-holes in an
algebraic way. In addition to the $R$-matrix, one will also need
to know the $6j$-symbols, because to describe general braidings,
on needs to know how to change between the different bases of
the tensor product of three representations. This information is
encoded in the ($q$-deformed) $6j$-symbols, or the $F$-matrices,
see for instance, \cite{pr} for a nice review.

To be more explicit, the $F$-matrices describe the basis transformation
between the two different ways in which one can take the tensor
product, or fusion, of three representations, or particle types.
The first way is to first fuse a particle of type $a$ with a particle of type $b$,
which gives, say, a particle of type $e$. Finally, one fuses this particle
$e$ which the third particle $c$, with particle $d$ as outcome.
The other way of fusing particles $a$, $b$ and $c$ is to first fuse
$b$ and $c$ into $f$, which is fused with $a$ to give $d$. Pictorially,
we can describe the relation between these two bases as
\begin{equation}
\label{fdef}
\psset{unit=.5mm,linewidth=.2mm,dimen=middle}
\begin{pspicture}(0,0)(105,35)
\psline(0,35)(0,30)(30,0)
\psline(10,35)(10,20)
\psline(20,35)(20,10)
\rput(3,32.5){$a$}
\rput(13,32.5){$b$}
\rput(23,32.5){$c$}
\rput(33,3){$d$}
\rput(12,13){$e$}
\rput(55,17.5){$= \sum (F^{a,b,c}_d)_{e,f}$}
\psline(70,35)(70,30)(100,0)
\psline(80,35)(80,30)(85,25)(90,30)(90,35)
\psline(85,25)(85,15)
\rput(73,32.5){$a$}
\rput(83,32.5){$b$}
\rput(93,32.5){$c$}
\rput(103,3){$d$}
\rput(87.5,20){$f$}
\end{pspicture} \ .
\end{equation}

The exchange of two particles $a$ and $b$ in a definite fusion channel
$c$ is described by the $R$-matrix. Pictorially, we have
\begin{equation}
\label{rdef}
\psset{unit=.5mm,linewidth=.2mm,dimen=middle}
\begin{pspicture}(-10,-5)(30,20)
\rput(-10,7.5){$R^{a,b}_c$}
\psline(0,20)(5,5)(10,20)
\psline(5,5)(5,-5)
\rput(-3,17.5){$a$}
\rput(12.5,17.5){$b$}
\rput(7.5,-2.5){$c$}
\rput(17.5,7.5){$=$}
\psline(35,20)(25,10)(30,5)(35,10)(31.4,13.6)
\psline(28.6,16.4)(25,20)
\psline(30,5)(30,-5)
\rput(22,17.5){$a$}
\rput(37.5,17.5){$b$}
\rput(32.5,-2.5){$c$}
\end{pspicture} \ .
\end{equation}

In order that the $F$ and $R$ matrices describe consistent braiding,
they have to satisfy consistency conditions, which go under the name
of the pentagon and hexagon equations \cite{ms1}. The $F$ and $R$ matrices
obtained from the quantum groups automatically satisfy these
equations.

We can now express the braiding of quasi-holes in terms of the $R$
and $F$-matrices. The braiding of the quasi-holes $1$ and $2$
(using the notation of sections \ref{rrsec} and \ref{sec:asstates}),
is simply given by the elements of the $R$ matrix, and depends on
the fusion channel.

The exchange of particles $2$ and $3$ can be done in more than
one way. In our case, it turns out that we first need to exhange
particles $1$ and $2$, followed by acting with the $F$-matrix,
and finally, exchanging the third particle with the `intermediate'
particle. Pictorially, we have (for clarity, we dropped the labels
$a$, $b$, etc.)
\begin{center}
\psset{unit=.5mm,linewidth=.2mm,dimen=middle}
\begin{pspicture}(0,0)(220,35)
\psline(0,35)(0,30)(30,0)
\psline(10,35)(10,20)
\psline(20,35)(20,10)
\psline[arrows=->](35,15)(45,15)
\rput(40,20){$R$}
\psline(50,35)(50,30)(55,25)(58,25)
\psline(62,25)(65,25)(65,15)
\psline(60,35)(60,20)(80,0)
\psline(70,35)(70,10)
\psline[arrows=->](85,15)(95,15)
\rput(90,22){$\sum F$}
\psline(100,35)(100,30)(105,25)(108,25)
\psline(112,25)(120,25)(120,10)
\psline(110,35)(110,20)(130,0)
\psline(120,35)(120,25)
\psline[arrows=->](135,15)(145,15)
\rput(140,20){$R$}
\psline(150,35)(150,30)(155,25)(158,25)
\psline(162,25)(170,25)(170,10)(180,0)
\psline(160,35)(160,20)(168,20)
\psline(172,20)(175,20)(175,5)
\psline(170,35)(170,25)
\rput(185,15){$=$}
\psline(190,35)(190,30)(198.6,21.4)
\psline(201.4,18.6)(220,0)
\psline(200,35)(200,10)(208,10)
\psline(212,10)(215,10)(215,5)
\psline(210,35)(210,20)(205,15)
\end{pspicture} \ .
\end{center}
This corresponds to the following form of the braid matrix $U_{23}$
\begin{equation}
\label{u23}
(U_{23})_{e,f} = R^{b,f}_d (F^{a,b,c}_{d})^e_f R^{a,b}_{e} \ . 
\end{equation}
Note that we do not sum over repeated indices. Similarly, we have the
following expression for $U_{13}$
\begin{equation}
\label{u13}
(U_{13})_{e,f} = R^{a,f}_{d} (R^{b,c}_f)^{-1} (F^{a,b,c}_{d})^e_f \ .
\end{equation} 
We will use the expressions \eqref{u23} and \eqref{u13} in the following
subsections to compare the braid results of the previous sections with the
results obtained using the quantum groups. Note that by changing the
relative phase of the different blocks, we can change the relative phase between
the off diagonal elements of the matrices $U_{13}$ and $U_{23}$. Thus, only the
sum of the off-diagonal phases is a physical quantity. Nevertheless, we reproduce
the matrices exactly.

We will not explain in detail how to obtain the $F$ and
$R$ matrices from the quantum groups. For this, we refer to \cite{sb1}
and a forthcoming paper \cite{afs}, in which details will be given
on the calculation of the $F$ and $R$ matrices in the case of
$U_q (su(3))$. Here, we will be brief in our description, and quote
the results we need from the literature.

A quantum group associated to a Lie algebra is a $q$-deformation of
its universal enveloping algebra. At generic values of $q$, the
representations of the quantum group are similar to the 
representations of the Lie algebra. However, if $q$ equals
specific roots of unity (which we will specify below in the cases of
our interest), the representation theory is rather different.
Concentration on the case $U_q (su(2))$ (see, for instance, \cite{sb1},
where the quantum group picture was used to calculate the
braid behaviour of the quasi-holes of the Read-Rezayi states),
we find that for $q=e^{\frac{2\pi i}{k+2}}$, there are $k+1$ unitary
highest weight representations. In addition, the tensor product of
such representations gets truncated in comparison to the
tensor product of $su(2)$. In fact, the truncated tensor products
are equivalent to the fusion rules of $su(2)_k$. The braid properties
of the quasi-holes are in fact described by the $F$ and $R$ matrices
of the quantum group at these special values of $q$.

The calculation of the $F$ matrices can be done in a similar fashion as
in the case of ordinary groups, by first calculating the Clebsch-Gordan
coefficients, and from those the $6j$-coefficients. Of course, in the
quantum group case, one has to use the $q$-deformed raising and
lowering operators.

Generically, the $F$-matrices can be expressed in terms of the
so-called $q$ numbers $\qnum{n}$, which are defined as
\begin{equation}
\qnum{n} = \frac{q^{\frac{n}{2}}-q^{-\frac{n}{2}}}{q^{\frac{1}{2}}-q^{-\frac{1}{2}}} \ ,
\end{equation}
or, equivalently, $\qnum{n} = \sum_{i=1}^{n} q^{\frac{n+1}{2}-i}$.
Note that for $q=1$, we have $\qnum{n}=n$.

For the Read-Rezayi states
with $k=3$ and the paired spin-singlet states proposed by the authors,
the corresponding value of $q$ is $q=e^{\frac{2 \pi i}{5}}$, for which we have
$\qnum{0}=0,\qnum{1} = \qnum{4} = 1,\qnum{2}=\qnum{3}=\frac{1+\sqrt{5}}{2}$ and
$\qnum{n+5} = -\qnum{n}$ for $n\in \NN$.

\subsection{The case $su(2)_k$}

To calculate the braid properties of the quasi-holes over the Read-Rezayi states, we
need to specify to which $su(2)_k$ representation they correspond, and the possible
fusion channels as well. Because all the braid properties are encoded by the braid
properties of four quasi-holes and because we know the four quasi-hole correlators,
we will focus on that case. The corresponding quantum group is
$U_q(su(2))$, with $q=e^{\frac{2 \pi i}{k+2}}$.
The quasi-holes correspond to the representation $l=1$ (i.e. spin $\frac{1}{2}$),
and the two fusion channels to the representations $l=0$ and $l=2$.

Let us focus on the matrix $U_{12}$ first. This matrix describes the braiding
of the first two quasi-holes, which depends on the fusion channel.
For $k=3$, we have
$R^{1,1}_{0} = (-1)^{\frac{7}{10}}$ and $R^{1,1}_{2} = (-1)^{\frac{1}{10}}$.
This is in agreement with the matrix $U_{12}$ of eq.~\eqref{rrk3braid}.

For general $k$, we have $R^{1,1}_0 = -q^{-\frac{3}{4}} = (-1)^{\frac{2k+1}{2(k+2)}}$ 
and $R^{1,1}_2 = q^{\frac{1}{4}} = (-1)^{\frac{1}{2(k+2)}}$, which confirms
$U_{12}$ of eq.~\eqref{rrgenbraid}. Note that the $M$ dependence easily follows
from the general form of the wavefunction, eq.~\eqref{rrkm}.

In addition, we have for $k=3$
\begin{equation}
\label{fmatsu23}
F^{1,1,1}_{1} = \begin{pmatrix}
-\tau & \sqrt{\tau} \\ \sqrt{\tau} & \tau \\
\end{pmatrix} \ .
\end{equation}
For general $k$, we obtain the following result
\begin{equation}
F^{1,1,1}_{1} = \frac{1}{\qnum{2}}\begin{pmatrix}
-1 & \sqrt{\qnum{3}} \\ \sqrt{\qnum{3}} & 1
\end{pmatrix} 
= \frac{1}{d_k} \begin{pmatrix}
-1 & \sqrt{d_k^2-1} \\ \sqrt{d_k^2-1} & 1 
\end{pmatrix} \ ,
\end{equation}
where $d_k = 2 \cos \bigl( \frac{\pi}{k+2}\bigr)$ is the quantum
dimension of the fundamental representation of $su(2)_k$.

Now, the matrix $U_{23}$ corresponds to
\begin{equation}
(U_{23})_{e,f} = R^{b,f}_d (F^{a,b,c}_{d})^e_f R^{a,b}_{e} \ , 
\end{equation} 
with $a=b=c=d=1$ and both $e$ and $f$ can take the values $0,2$.
Note that we do not sum over repeated indices. Similarly, $U_{13}$ corresponds 
to
\begin{equation}
(U_{13})_{e,f} = R^{a,f}_{d} (R^{b,c}_f)^{-1} (F^{a,b,c}_{d})^e_f \ , 
\end{equation} 
also with $a=b=c=d=1$ and $e,f=0,2$.

Using these results for the $R$ and $F$ matrices, we easily see that the
$k=3$ results of eq.~\eqref{rrk3braid} and the general results of
eq.~\eqref{rrgenbraid} are reproduced exactly.

We would like to note that the $k=3$ braid matrices are, up to an overall 
factor, directly related to the braid matrices of the Fibonacci theory. 
Note that they are in fact related to the `mirror-image' of the matrices 
\eqref{eq:RFB}. This is related to the fact that the $F$ matrix of the 
spin-$1/2$ particles in the $su(2)_3$ theory is given by eq.~\eqref{fmatsu23} 
instead of eq.~\eqref{eq:RFB}.

\subsection{The case $su(3)_2$}

We will now compare the results of braiding in the spin-singlet case to 
the results which can be obtained by using a quantum group picture. To do 
this, we will have to know the $6j$-symbols (or $F$-symbols) of the quantum 
group of $su(3)$. The results of a direct calculation of these $6j$-symbols 
for general $q$ will be presented elsewhere \cite{afs}, together with 
the $R$-matrix `eigenvalues' and a detailed analysis of the $6j$-symbols 
related to cosets. Here, we will merely quote a small number of $F$-symbols 
and braid factors.

For the comparison of the braid matrices with the results obtained
from the quantum group picture, it is easiest to work with a
set of parafermion fields which stay as close as possible to the
representations used in the KZ equation, see appendix \ref{sec:su32pf}.
Hence, we will work with the representatives $a={\bf 3}$, $b={\bf 3}$, 
$c={\bf 3}$, and $d={\bf 8}$. Note that we can not take ${\bf 3}$ as the 
last representative
as well. The reason is that after fusing the first two fields, we find that
the possible channels correspond to ${\bf 6}$ or ${\bf \bar{3}}$.
Fusing these intermediate channels automatically gives us the
${\bf 8}$. It follows that we need the following data, which can be
obtained from the quantum group of $su(3)$ \cite{afs}

\begin{equation}
\label{fmatsu32}
F^{{\bf 3},{\bf 3},{\bf 3}}_{\bf 8} = \begin{pmatrix}
\tau & \sqrt{\tau} \\ \sqrt{\tau} & -\tau \\
\end{pmatrix} \ ,
\end{equation}
where the intermediate fusion channels are ${\bf 6}$ and ${\bf \bar{3}}$ 
(in that order) and
\begin{align}
\label{rsu32}
R^{{\bf 3},{\bf 3}}_{\bf 6} &= (-1)^{\frac{2}{15}}  &
R^{{\bf 3},{\bf 3}}_{{\bf \bar{3}}} &= (-1)^{\frac{11}{15}}  &
R^{{\bf 3},{\bf 6}}_{\bf 8} &= (-1)^{\frac{2}{3}}  &
R^{{\bf 3},{\bf \bar{3}}}_{\bf 8} &= (-1)^{\frac{1}{15}} \ .  
\end{align}
The symbols 
$R^{{\bf 3},{\bf 3}}_{\bf 6}$ and $R^{{\bf 3},{\bf 3}}_{\bf \bar{3}}$
correspond to the diagonal elements of $U_{12}$ as they should.
In addition, upon using equations
\eqref{u23} and \eqref{u13}, we obtain the matrices of 
eq.~\eqref{eq:braiduuud}.

\subsection{$su(3)_2$ parafermion correlators}

To verify the braid behaviour of the various $su(3)_2$ parafermion correlators,
we will use the following data
\begin{equation}
F^{{\bf 8},{\bf 8},{\bf 8}}_{{\bf 8}} = \begin{pmatrix}
\tau & \sqrt{\tau} \\ \sqrt{\tau} & -\tau \\
\end{pmatrix} \ ,
\end{equation}
where the intermediate fusion channels are $\id$ and ${\bf 8}$ and
\begin{align}
R^{{\bf 8},\id}_{\bf 8} &= R^{\id,{\bf 8}}_{\bf 8} = 1 &
R^{{\bf 8},{\bf 8}}_{\id} & = (-1)^{-\frac{6}{5}} &
R^{{\bf 8},{\bf 8}}_{\bf 8} & = (-1)^{-\frac{3}{5}} \ .
\end{align}
With this data, we exactly obtain the braid behaviour which can be derived
from the correlator of four $\bar{\rho}_1$ fields in eq. \eqref{eq:fourrho}.
In addition, we obtain the braid behaviour of four $\sigma_3$ fields of
eq. \eqref{eq:fourspin} up to an overall sign. To explain the origin of this
sign, we note that if one expresses the $SU(3)_2$ WZW primary
in the adjoint representation in terms of the parafermion field
$\sigma_3$, additional $u(1)$ factors are needed, see
eq.~\eqref{su32pf}. These $u(1)$ factors give rise to the additional
sign. Note that in the $\rho$ sector, these $u(1)$ factors are absent.

\vskip 1cm

\appendix
\section{Detailed structure of the $su(2)_k$ parafermion theory}
\label{su2app}

In this appendix, we will review the structure of the $\ZZ_k$ parafermion
conformal field theory, as introduced in \cite{fz1}. In fact, we will use the
general setting of \cite{gep1}, so we can use the results to explain the
details of the parafermion theory associated to $su(3)_2$ in the next
appendix as well.

In the case of the $su(3)_2$ parafermions, which we will describe in the
next appendix, it will be important to make a distinction between the
chiral sectors of the parafermion theory, and the Virasoro primary fields.
Each chiral sector contains an infinite number of Virasoro
primary fields. However, we will often use the same notation for the
chiral sector and the leading Virasoro primary field. For the chiral
sectors we consider in this appendix, there will always be only one
leading virasoro primary field. However, in the case of the $su(3)_2$ 
parafermions, there will be one chiral sector containing two leading 
Virasoro primary fields.

The fusion rules (which will be specified for the $su(2)_k$ parafermions below)
describe the `merging' of two parafermion sectors. To calculate the quasi-hole
wavefunctions, we need to know the full details of what happens when two
primary fields are brought to the same location inside correlators. This 
information is contained in the Operator Product Expansions (OPE's), which 
will be given in \ref{aope} in the case of the parafermions associated to 
$su(2)_k$.

\subsection{Fusion rules of the $su(2)_3$ parafermions}
\label{afus}

The sectors of the $su(2)_k/u(1)$ parafermion CFT (which we will sometimes
denote by $\ZZ_k$) are labeled by two
labels, an $su(2)_k$ label $l=0,1,\ldots,k$ and the $u(1)$ charge $m$,
which is defined modulo $2k$, because the $u(1)$ theory is compactified.
Thus, we write the sector (and the leading parafermion primary fields)
as $\Phi^l_m$.

The branching rules state that the only labels allowed are those which
satisfy $l-m = 0 \mmod 2$. In addition, we need to identify the sectors \cite{gep2}
$\Phi^l_m \equiv \Phi^{k-l}_{m+k}$. It follows that there are $\frac{1}{2}k (k+1)$
parafermion primary fields, and for each field we can choose labels
$\Phi^l_m$ with $l=0,1,\ldots k$ and $m\in \{-l+2,-l+4,\ldots, l\}$. With labels
chosen in this way, the dimensions of the fields are given by
\begin{equation}
\label{hsu2kpf}
h_{l,m} = \frac{l(l+2)}{4(k+2)} - \frac{m^2}{4k} \ .
\end{equation}
Note that if $m$ is chosen outside of the range $m\in \{-l+2,-l+4,\ldots, l\}$, the
scaling dimension will be given by $h_{l,m} + n^l_m$, where $n^l_m$ is a
positive integer.

Thus, for the parafermions $\psi_i = \Phi^{0}_{2i}\equiv \Phi^{k}_{2i-k}$, with
$i=0,1,\ldots,k-1$ we find $h_{\psi_i} = \frac{i(k-i)}{k}$, while for the spin fields
$\sigma_i = \Phi^{i}_{i}$ we find $h_{\sigma_i} = \frac{i(k-i)}{2k(k+2)}$. The
`neutral' fields $\vep_i = \Phi^{2i}_{0}$ with $2i\in \{0,1,\ldots,k\}$
have scaling dimension $h_{\vep_i} = \frac{i(i+1)}{k+2}$.

The fusion rules of the $\ZZ_k$ parafermion theory can be obtained from
the fusion rules of the $SU(2)_k$ WZW conformal field theory. Explicitly, we
have
\begin{equation}
\label{su2kpffr}
\Phi^{l}_{m} \times \Phi^{l'}_{m'} = \oplus_{l''\in l\times l'} \Phi^{l''}_{m+m'} \ , 
\end{equation}
where the sum is over the range $l'' = |l-l'|, |l-l'|+2 ,\ldots, \min(l+l',2k-l-l')$.
 
Specializing to the $\ZZ_3$ parafermions, we will use the following
standard notation
\begin{align}
\id &= \Phi^0_0 &
\psi_1 &= \Phi^0_2 &
\psi_2 &= \Phi^0_4 &
\sigma_1 &= \Phi^1_1 &
\sigma_2 &= \Phi^2_2 &
\varepsilon &= \Phi^2_0 \ .
\end{align}
The scaling dimensions are $h_{\psi_i} = \frac{2}{3}$,
$h_{\sigma_i}=\frac{1}{15}$ and $h_\vep = \frac{2}{5}$. 

With this notation, we find the fusion rules as given in table \ref{fusrulsu23}.
\begin{table}[ht]
\begin{center}
\begin{tabular}{|c|c|c|c|c|c|}
\hline
$\times$ & $\sigma_1$ & $\sigma_2$ &
$\varepsilon$  & $\psi_1$ & $\psi_2$
\\ \hline
$\sigma_1$ & $\psi_1 + \sigma_2$ &&&&\\
$\sigma_2$ & $\id +\varepsilon$  & $\psi_2 + \sigma_1$ &&&\\
$\varepsilon$ & $\psi_2 + \sigma_1$ & $\psi_1 + \sigma_2$ &
$\id + \varepsilon$ & &\\
$\psi_1$ & $\varepsilon$ & $\sigma_1$ & $\sigma_2$ & $\psi_2$ &\\
$\psi_2$ & $\sigma_2$ & $\varepsilon$ & $\sigma_1$ & $\id$ & $\psi_1$\\ 
\hline
\end{tabular}
\end{center}
\vskip 2mm \caption{Fusion rules of the parafermion and spin
fields associated to the parafermion theory $su(2)_3/u(1)$.} 
\label{fusrulsu23}
\end{table}

Note that the structure of the fusion rules becomes simpler if we adopt the
following notation $\psi_0=\id$, $\psi_1 = \psi_1$, $\psi_2=\psi_2$,
$\tau_0=\varepsilon$, $\tau_1 =\sigma_2$ 
and $\tau_2=\sigma_1$. The labels of the fields are defined modulo 3.
In this notation, the fusion rules are simply
\begin{equation}
\begin{split}
\psi_i \times \psi_j &= \psi_{i+j} \\
\psi_i \times \tau_j &= \tau_{i+j} \\
\tau_i \times \tau_j &= \psi_{i+j} + \tau_{i+j}
\end{split}
\end{equation}

\subsection{OPE's}
\label{aope}

In this section, we give the OPE's of the leading Virasoro
primary fields%
\footnote{To be completely rigorous, we should view the
fields as chiral vertex operators, or intertwiners. Then, the
OPE coefficients would carry two additional labels, indicating
the sectors the fields acts between. In this paper, we will not need
this level of detail.}
(including the numerical coefficients). We will use 
schematic notation, writing
\begin{equation}
A(z)B(w)=(z-w)^{h_C-h_A-h_B} c_{A,B}^{C}  C(w) + \ldots
\label{eq:OPEAB1}
\end{equation}
as
\begin{equation}
AB=c_{A,B}^C C
\label{eq:OPEAB2}
\end{equation}
and restricting ourselves to the leading field in each 
of the fusion channels.

We then have the following OPE's
\begin{align}
\psi_1 \psi_1 &= \frac{2}{\sqrt{3}} \psi_2&
\psi_1 \psi_2 &= \id &
\psi_2 \psi_2 &= \frac{2}{\sqrt{3}} \psi_1 \ ,
\end{align}
in accordance with the results in \cite{fz1} for all fields 
$\psi_i$ and general $k$.
The OPE's between parafermions and spin fields are
\begin{align}
\psi_1 \sigma_1 &= \sqrt{\frac{2}{3}} \varepsilon &
\psi_1 \sigma_2 &= \frac{1}{\sqrt{3}} \sigma_1 &
\psi_1 \varepsilon &= \frac{2}{\sqrt{3}} \sigma_2 
\nonu
\psi_2 \sigma_1 &= \frac{1}{\sqrt{3}} \sigma_2 &
\psi_2 \sigma_2 &= \sqrt{\frac{2}{3}} \varepsilon &
\psi_2 \varepsilon &= \frac{2}{\sqrt{3}} \sigma_1 \ ,
\end{align}
while the other OPE's of the spin fields are given by,
with $C = \frac{1}{2}\sqrt{
\frac{\Gamma(\frac{1}{5})\Gamma^3(\frac{3}{5})}
{\Gamma(\frac{4}{5})\Gamma^3(\frac{2}{5})}}$
\bea
&& 
\sigma_1 \sigma_1 = \frac{1}{\sqrt{3}} \psi_1 + \sqrt{2C} \sigma_2 
\qquad
\sigma_1 \sigma_2 = \id + \sqrt{C} \varepsilon 
\qquad
\sigma_2 \sigma_2 = \frac{1}{\sqrt{3}} \psi_2 + \sqrt{2C} \sigma_1 
\nonu 
&& \qquad\qquad 
\sigma_1 \varepsilon = \sqrt{\frac{2}{3}} \psi_2 + \sqrt{C} \sigma_1
\qquad
\sigma_2 \varepsilon = \sqrt{\frac{2}{3}} \psi_1 + \sqrt{C} \sigma_2 \ .
\eea
The most interesting OPE turns out to be
\begin{equation}
\varepsilon \varepsilon =
\id + 0\,  \varepsilon + \sqrt{\frac{12 C}{7}} \varepsilon' \ ,
\end{equation} 
that is, the first field appearing in the `$\varepsilon$' channel is {\em not}
the field $\varepsilon$, but a different Virasoro primary field $\varepsilon'$,
which has scaling dimension $h_{\varepsilon'} = \frac{7}{5}$. This result
will be derived in section \ref{sec:morecorrsu23}.

\subsection{Spin field correlators}
\label{scor}

In this appendix, we will explain how the four point correlators
of four `spin fields' $\sigma$ can be obtained from the
four point correlators of the WZW CFT as given in \cite{kz1}.
We will do this explicitly in the case of $su(2)_3$, and merely state the
outcome in the other cases. 

The correlators which are calculated in \cite{kz1} are four point
correlators of WZW primary fields, transforming in the fundamental
representation. These fields $g$ can be written in terms of the
spin fields $\sigma_1$ and $\sigma_2$, combined with a vertex operator.
Explicitly, we have $g_1 = \sigma_1 e^{\frac{i \varphi}{6}}$ and
$g_2 = \sigma_2 e^{-\frac{i \varphi}{6}}$. Because correlators of the form
$\left< g_1 g_2 g_1 g_2 \right>^{(p)}$ and
$\left< g_1 g_1 g_2 g_2 \right>^{(p)}$ are given in \cite{kz1} and four
point correlators of the vertex operators $e^{\pm {\frac{i \varphi}{6}}}$
are easily calculated, we can derive the explicit form of the correlators
$\left<\sigma_1 \sigma_2 \sigma_1 \sigma_2 \right>^{(p)}$ and
$\left< \sigma_1 \sigma_1 \sigma_2 \sigma_2 \right>^{(p)}$, namely
\begin{align}
\left<\sigma_1 (w_1)\sigma_2 (w_2)\sigma_1 (w_3)\sigma_2 (w_4)\right>^{(0)} 
&= (w_{12}w_{34})^{-\frac{2}{15}} x^{\frac{3}{10}} 
(1-x)^{-\frac{1}{6}} \cF^{(0)}_1 (x) 
\nonu
\left<\sigma_1 (w_1)\sigma_2 (w_2)\sigma_1 (w_3)\sigma_2 (w_4)\right>^{(1)} &= 
(-1)^{\frac{2}{5}} (w_{12}w_{34})^{-\frac{2}{15}} x^{\frac{3}{10}} 
(1-x)^{-\frac{1}{6}} C \cF^{(1)}_1 (x) 
\nonu 
\left<\sigma_1 (w_1)\sigma_1 (w_2)\sigma_2 (w_3)\sigma_2 (w_4)\right>^{(0)} &= 
(-1)^{\frac{2}{3}} (w_{12}w_{34})^{-\frac{2}{15}} x^{-\frac{1}{30}} 
(1-x)^{\frac{1}{6}} \cF^{(0)}_2(x) 
\nonu
\left<\sigma_1 (w_1)\sigma_1 (w_2)\sigma_2 (w_3)\sigma_2 (w_4)\right>^{(1)} &= 
-(-1)^{\frac{1}{15}}(w_{12}w_{34})^{-\frac{2}{15}} x^{-\frac{1}{30}} 
(1-x)^{\frac{1}{6}} C \cF^{(1)}_2 (x) \ .
\end{align}
The constant $C$ is related to $h$ used in \cite{kz1} via $h=C^2$.
The phases make sure that the correlators have the correct behaviour when 
$x\to 0$. In this limit, the four point correlators reduce to normalized 
two point functions. We remind the reader of the notation $w_{ij} = w_i-w_j$. 
The functions $\cF^{(p)}_i(x)$ can be expressed in terms of the hypergeometric
functions $F(a,b,c;x)$ in the following way
\begin{align}
\cF^{(0)}_1 (x) &= x^{-\frac{3}{10}} (1-x)^{\frac{1}{10}}
F(\frac{1}{5},-\frac{1}{5},\frac{3}{5};x) &
\cF^{(0)}_2 (x) &= \frac{1}{3} x^{\frac{7}{10}} (1-x)^{\frac{1}{10}}
F(\frac{6}{5},\frac{4}{5},\frac{8}{5};x) 
\nonu
\cF^{(1)}_1 (x) &= x^{\frac{1}{10}} (1-x)^{\frac{1}{10}}
F(\frac{1}{5},\frac{3}{5},\frac{7}{5};x)  &
\cF^{(1)}_2 (x) &= -2 x^{\frac{1}{10}} (1-x)^{\frac{1}{10}}
F(\frac{1}{5},\frac{3}{5},\frac{2}{5};x) \ .
\label{fsu23}
\end{align}

\subsection{Further correlators}
\label{sec:morecorrsu23}

The master formula \eqref{meq} can be used to obtain other
correlation functions of parafermion fields, which are hard to
obtain from the KZ equation, because they correspond to
other SU(2) representations than the fundamental representations.

These correlators are obtained by taking various limits of the master
formula \eqref{meq}. In fact, for $k=3$, we can obtain all the possible
four point correlation functions of the spin fields $\sigma_1$,$\sigma_2$
and $\varepsilon$, because we can fuse the spin field $\sigma_1$ with
an arbitrary number of fields $\psi_i$, which gives us all the possible
spin fields. For arbitrary $k$, these methods gives us a way of
calculating all the four point correlators of the fields of the form
$\Phi^{1}_m \equiv \Phi^{k-1}_{m+k}$, in combination with an arbitrary
number (which has to be allowed by fusion) of parafermion fields $\psi_i$. 

We will now use the master formula \eqref{meq} to find the correlator\\
$\left<\sigma_1 (w_1)\sigma_2 (w_2) \vep (w_3) \vep (w_4)\right>^{(0,1)}$,
by taking the limit
($z_1\rightarrow w_2$,$z_2\rightarrow w_2$,$z_3\rightarrow w_3$,
$z_4\rightarrow w_4$).
This results in
\begin{align}
\left<\sigma_1 (w_1)\sigma_2 (w_2) \vep (w_3) \vep (w_4)\right>^{(0)} &=
\frac{1}{2} w_{12}^{-\frac{2}{15}} w_{34}^{-\frac{4}{5}} x^{\frac{3}{10}} (1-
x)^{\frac{1}{2}}
\Bigl[\frac{2-x}{1-x} \cF^{(0)}_1(x)+\cF^{(0)}_2(x)\Bigr] 
\\[2mm]
\left<\sigma_1 (w_1)\sigma_2 (w_2)\vep (w_3)\vep (w_4)\right>^{(1)} &= 
(-1)^{\frac{7}{5}}
\frac{C}{2} w_{12}^{-\frac{2}{15}} w_{34}^{-\frac{4}{5}} x^{\frac{3}{10}} 
(1-x)^{\frac{1}{2}}
\Bigl[\frac{2-x}{1-x} \cF^{(1)}_1(x)+\cF^{(1)}_2(x)\Bigr] 
\nonumber
\end{align}
Here, we used that $C_{\psi_1,\sigma_1}^{\vep}= \sqrt{\frac{2}{3}}$. We 
would like to note
that this result is equivalent to the result obtained in \cite{fz1}, 
namely
\begin{align}
\left<\sigma_1 (w_1)\sigma_2 (w_2)\vep (w_3)\vep (w_4)\right>^{(0)} &=
w_{12}^{-\frac{2}{15}} w_{34}^{-\frac{4}{5}} (1-x)^{-\frac{2}{5}}
F(-\frac{1}{5},-\frac{4}{5},-\frac{2}{5};x) 
\nonu
\left<\sigma_1 (w_1)\sigma_2 (w_2)\vep (w_3)\vep (w_4)\right>^{(1)} &= 
(-1)^{\frac{7}{5}}
w_{12}^{-\frac{2}{15}} w_{34}^{-\frac{4}{5}} x^{\frac{7}{5}} (1-x)^{-
\frac{2}{5}} \sqrt{\rho}
F(\frac{6}{5},\frac{3}{5},\frac{12}{5};x) \ .
\end{align}
The equivalence of the two results follows from the fact that 
$\sqrt{\rho}/C=\frac{2}{7}$, with $\rho$ given by
\begin{equation}
\rho = 4 \frac{\Gamma^3(\frac{3}{5})\Gamma^2(\frac{6}{5})\Gamma(\frac{4}{5})}
{\Gamma^2(\frac{12}{5})\Gamma^2(-
\frac{1}{5})\Gamma(\frac{2}{5})\Gamma(\frac{1}{5})}
\end{equation}
and the following relations between hypergeometric functions
\begin{align}
(1-x)^{-\frac{2}{5}} F(-\frac{1}{5},-\frac{4}{5},-\frac{2}{5};x) &=
\frac{1}{2} x^{\frac{3}{10}} (1-x)^{\frac{1}{2}} \Bigl[\frac{2-x}{1-x} 
\cF^{(0)}_1(x)+\cF^{(0)}_2(x)\Bigr] 
\nonu
x^{\frac{7}{5}}(1-x)^{-\frac{2}{5}} F(\frac{6}{5},\frac{3}{5},\frac{12}{5};x) 
&= \frac{7}{4} x^{\frac{3}{10}} (1-x)^{\frac{1}{2}} \Bigl[\frac{2-x}{1-x} 
\cF^{(1)}_1(x)+\cF^{(1)}_2(x)\Bigr] 
\end{align}
To obtain the OPE coefficient $C_{\vep,\vep}^\vep$, we expand the
correlators around $x=0$, with the result
\begin{align}
w_{12}^{\frac{2}{15}} w_{34}^{\frac{4}{5}}
\left<\sigma_1 (w_1)\sigma_2 (w_2)\vep (w_3)\vep (w_4)\right>^{(0)} &=
1+ \frac{x^2}{15} + \frac{x^3}{15} + o(x^4) 
\nonu
w_{12}^{\frac{2}{15}} w_{34}^{\frac{4}{5}} \frac{(-1)^{-\frac{7}{5}}7}{2 
\sqrt{h}}
\left<\sigma_1 (w_1)\sigma_2 (w_2)\vep (w_3)\vep (w_4)\right>^{(1)} &=
x^{\frac{7}{5}} + \frac{7 x^{\frac{12}{5}}}{10} + \frac{236 
x^{\frac{17}{5}}}{425} + o(x^{\frac{22}{5}}) 
\end{align}
We find that the $1$ (or $\sigma_0)$ channel starts as 
$x^{\frac{7}{5}}$, as was observed in \cite{fz1}. As a consequence, we
find that the OPE coefficient $C_{\vep,\vep}^\vep = 0$.

Finally, we also reduced the master formula \eqref{meq} to obtain
the correlator \\
$\left<\vep (w_1)\vep (w_2)\vep (w_3)\vep (w_4)\right>^{(0,1)}$,
by taking the limit
($z_1\rightarrow w_2$,$z_2\rightarrow w_1$,$z_3\rightarrow w_3$,$z_4
\rightarrow w_4$).
This results in
\begin{align}
\left<\vep (w_1)\vep (w_2)\vep (w_3)\vep (w_4)\right>^{(0)} &=
\frac{1}{2} (w_{12}w_{34})^{-\frac{4}{5}} x^{\frac{3}{10}} (1-x)^{-\frac{1}{2}}
[(2-x) \cF^{(0)}_1(x)+(1+x)\cF^{(0)}_2(x)]
\nonu
\left<\vep (w_1)\vep (w_2)\vep (w_3)\vep (w_4)\right>^{(1)} &= 
\frac{-(-1)^{\frac{7}{5}}C}{2 (w_{12}w_{34})^{\frac{4}{5}}} 
x^{\frac{3}{10}} (1-x)^{-\frac{1}{2}}
[(2-x) \cF^{(1)}_1(x)+(1+x)\cF^{(1)}_2(x)]
\end{align}
Again, we find that the small $x$ behaviour for the $(1)$ or $\vep$ channel 
goes like $x^{\frac{7}{5}}$, explicitly
\begin{align}
(w_{12}w_{34})^{\frac{4}{5}}
\left<\vep (w_1)\vep (w_2)\vep (w_3)\vep (w_4)\right>^{(0)} &=
1+ \frac{2 x^2}{5} + \frac{2 x^3}{5} + o(x^4) 
\nonu
(w_{12}w_{34})^{\frac{4}{5}} \frac{(-1)^{-\frac{7}{5}}7}{12 C}
\left<\vep (w_1)\vep (w_2)\vep (w_3)\vep (w_4)\right>^{(1)} &=
x^{\frac{7}{5}} + \frac{7 x^{\frac{12}{5}}}{10} + \frac{261 
x^{\frac{17}{5}}}{425} + o(x^{\frac{22}{5}}) 
\end{align}

Using the braid properties of the functions $\cF^{(0,1)}_{1,2}$, we find the
following braid matrices for the correlator 
$\left<\vep (w_1)\vep (w_2)\vep (w_3)\vep (w_4)\right>^{(0,1)}$
\begin{align}
U_{12} &= \begin{pmatrix}
(-1)^{-\frac{4}{5}} & 0 \\ 0 & (-1)^{\frac{3}{5}}
\end{pmatrix} & 
U_{23} &= (-1)^{-\frac{4}{5}} \begin{pmatrix}
\tau & (-1)^{-\frac{7}{5}}\sqrt{\tau} \\ (-1)^{\frac{7}{5}}\sqrt{\tau} & - \tau
\end{pmatrix} \nonu
U_{13} &= (-1)^{\frac{4}{5}} \begin{pmatrix}
\tau & (-1)^{-\frac{4}{5}}\sqrt{\tau} \\ \sqrt{\tau} & (-1)^{\frac{1}{5}} \tau
\end{pmatrix} \ .
\label{epbraid}
\end{align}
Again, we would like to see if
we can obtain the same result using the quantum group picture.
This time, we need the $R$ and $F$-matrices corresponding to
$l=2$ or spin-$1$, because the $\vep$ particles are represented
by $\vep = \Phi^2_0$  in the coset construction. The data we need
is $R^{2,2}_0=(-1)^{-\frac{4}{5}}$, $R^{2,2}_0=(-1)^{\frac{3}{5}}$
and
\be
F^{2,2,2}_2 = \begin{pmatrix}
\tau & -\sqrt{\tau} \\ -\sqrt{\tau} & -\tau 
\end{pmatrix} \ .
\ee
With this data, we can calculate the braid matrices using
\eqref{u23} and \eqref{u13}, with the following
result
\begin{align}
U_{12} &= \begin{pmatrix}
(-1)^{-\frac{4}{5}} & 0 \\ 0 & (-1)^{\frac{3}{5}}
\end{pmatrix} 
&
U_{23} &= (-1)^{-\frac{4}{5}} \begin{pmatrix} 
\tau & (-1)^{-\frac{2}{5}} \sqrt{\tau} \\ 
(-1)^{\frac{2}{5}}\sqrt{\tau} & - \tau
\end{pmatrix} 
\nonu
U_{13} &= (-1)^{\frac{4}{5}} \begin{pmatrix}
\tau & (-1)^{\frac{1}{5}} \sqrt{\tau} \\ 
-\sqrt{\tau} & (-1)^{\frac{1}{5}} \tau
\end{pmatrix}
\end{align}
We see that we almost exactly reproduced the braid behaviour of
eq.~\eqref{epbraid}. The only difference is the additional minus signs in the
off-diagonal elements of the matrices. These additional minus signs are
a result of the fact that the conformal block in the $(1)$ or $(\vep)$
channel starts at a degree higher than naively expected from the
fusion rules. However, these additional signs can be `gauged' away,
by redefining the block $\langle \vep\vep\vep\vep\rangle^{(1)}$ with
an additional sign. Thus, we conclude that the vanishing of the
coefficient $C_{\vep,\vep}^{\vep}$ does show up in the braid
matrices, but its effect can be gauged away. Note that is was to be
expected that the additional power in the correlator would not
change the braid behaviour, because there is no (or hardly any)
freedom for change without violating the pentagon and hexagon
equations.

\subsection{Braiding relations}
\label{su2br}

{}From \cite{kz1} we know that
\be
\cF_{1,2}^{(p)} (1-x) = \sum_q C^p_q \cF^{(q)}_{2,1} (x) \ ,
\ee
where, in this case, we have
\begin{align}
C^0_0 &= - C^1_1 =
\frac{\Gamma(\frac{2}{5})\Gamma(\frac{3}{5})}{\Gamma(\frac{1}{5})
\Gamma(\frac{4}{5})} = \frac{\sqrt{5}-1}{2}=\tau 
\nonu
C^1_0 &= -2 \frac{\Gamma^2(\frac{2}{5})}{\Gamma(\frac{1}{5})
\Gamma(\frac{3}{5})} \qquad\qquad
C^0_1 = \frac{1+C^0_0 C^1_1}{C^1_0} =
-\frac{1}{2} \frac{\Gamma^2(\frac{3}{5})}{\Gamma(\frac{2}{5})
\Gamma(\frac{4}{5})}
\end{align}
In addition, we have the following expression for $C$, which is related to
$h$ used in \cite{kz1}
\begin{align}
h &= C^2 = \frac{C^0_1}{C^1_0} = \frac{1}{4}
\frac{\Gamma(\frac{1}{5})\Gamma^3(\frac{3}{5})}{\Gamma(\frac{4}{5})\Gamma^3(\frac{2}{5})}
& \frac{C^0_1}{C} &= C^1_0 C = -\sqrt{\tau} \ .
\end{align}

For general $k$, these relations become
\begin{align}
C^0_0 &= - C^1_1 =
\frac{\Gamma(\frac{2}{k+2})\Gamma(\frac{k}{k+2})}{\Gamma(\frac{1}{k+2})
\Gamma(\frac{k+1}{k+2})} =
\frac{1}{2 \cos (\frac{\pi}{k+2})} 
\nonu
C^1_0 &= -2 
\frac{\Gamma^2(\frac{2}{k+2})}{\Gamma(\frac{1}{k+2})\Gamma(\frac{3}{k+2})} 
\qquad\qquad
C^0_1 = \frac{1+C^0_0 C^1_1}{C^1_0} =
-\frac{1}{2} \frac{\Gamma^2(\frac{k}{k+2})}{\Gamma(\frac{k-
1}{k+2})\Gamma(\frac{k+1}{k+2})}
\end{align}
In addition, we have the following expression for $h$, see \cite{kz1}
\begin{align}
h &= C^2 = \frac{C^0_1}{C^1_0} = \frac{1}{4}
\frac{\Gamma(\frac{1}{k+2})\Gamma(\frac{3}{k+2})\Gamma^2(\frac{k}{k+2})}
{\Gamma(\frac{k-1}{k+2})\Gamma(\frac{k+1}{k+2})\Gamma^2(\frac{2}{k+2})}
& \frac{C^0_1}{C} &= C^1_0 C =  -\sqrt{1+C^0_0 C^1_1} \ .
\end{align}

\noindent
The transformation behaviour of the $\cF^{(p)}_i (x)$ under
$x \mapsto \frac{-x}{1-x}$ is as follows
\begin{align}
\cF^{(0)}_1 (\frac{-x}{1-x}) & =
(-1)^{-\frac{3}{2(k+2)}} (1-x)^{\frac{3}{2(k+2)}} [ \cF^{(0)}_1 (x) + 
\cF^{(0)}_2 (x) ] \nonu
\cF^{(0)}_2 (\frac{-x}{1-x}) & =
-(-1)^{-\frac{3}{2(k+2)}} (1-x)^{\frac{3}{2(k+2)}} \cF^{(0)}_2 (x)\nonu
\cF^{(1)}_1 (\frac{-x}{1-x}) & =
-(-1)^{\frac{1}{2(k+2)}} (1-x)^{\frac{3}{2(k+2)}} [ \cF^{(1)}_1 (x) + 
\cF^{(1)}_2 (x) ] \nonu
\cF^{(1)}_2 (\frac{-x}{1-x}) & =
(-1)^{\frac{1}{2(k+2)}} (1-x)^{\frac{3}{2(k+2)}} \cF^{(1)}_2 (x) \ .
\end{align}

\noindent
For $x\mapsto \frac{1}{x}$ we have
\begin{align}
\cF^{(0)}_1 (\frac{1}{x}) &= (-1)^{\frac{3}{2(k+2)}} x^{\frac{3}{2(k+2)}}
[ C^0_0 \cF^{(0)}_1 (x) 
+ (-1)^{\frac{k}{k+2}} C^0_1 \cF^{(1)}_1 (x) ] 
\\[2mm]
\cF^{(0)}_2 (\frac{1}{x}) &= -(-1)^{\frac{3}{2(k+2)}} x^{\frac{3}{2(k+2)}}
[ C^0_0 \bigl( \cF^{(0)}_1 (x) + \cF^{(0)}_2 (x) \bigr) + 
(-1)^{\frac{k}{k+2}} C^0_1 \bigl( \cF^{(1)}_1 (x) + \cF^{(1)}_2 (x) \bigr)
] \nonu
\cF^{(1)}_1 (\frac{1}{x}) &= -(-1)^{-\frac{1}{2(k+2)}} x^{\frac{3}{2(k+2)}}
[ C^1_0 \cF^{(0)}_1 (x) + (-1)^{\frac{k}{k+2}} C^1_1 \cF^{(1)}_1 (x) ] 
\nonu
\cF^{(1)}_2 (\frac{1}{x}) &= (-1)^{-\frac{1}{2(k+2)}} x^{\frac{3}{2(k+2)}}
[ C^1_0 \bigl( \cF^{(0)}_1 (x) + \cF^{(0)}_2 (x) \bigr) + 
(-1)^{\frac{k}{k+2}} C^1_1 \bigl( \cF^{(1)}_1 (x) + \cF^{(1)}_2 (x) \bigr) 
] \ . \nonumber
\end{align}

\subsection{More general $k$ results}
\label{sec:rrk}

{}From \cite{kz1}, we can extract the following correlators for general $k$
\begin{align}
\left<\sigma_1 (w_1)\sigma_{k-1} (w_2)\sigma_1 (w_3)\sigma_{k-1} 
(w_4)\right>^{(0)} &=
(w_{12}w_{34})^{-\frac{k-1}{k(k+2)}}
x^{\frac{3}{2(k+2)}} (1-x)^{-\frac{1}{2k}} \cF^{(0)}_1 (x) 
\\[2mm]
\left<\sigma_1 (w_1)\sigma_{k-1} (w_2)\sigma_1 (w_3)\sigma_{k-1} 
(w_4)\right>^{(1)} &= 
(-1)^{\frac{2}{k+2}} (w_{12}w_{34})^{-\frac{k-1}{k(k+2)}}
x^{\frac{3}{2(k+2)}} (1-x)^{-\frac{1}{2k}} C \cF^{(1)}_1 (x) \nonu
\left<\sigma_1 (w_1)\sigma_1 (w_2)\sigma_{k-1} (w_3)\sigma_{k-1} 
(w_4)\right>^{(0)} &= 
(-1)^{\frac{k-1}{k}} (w_{12}w_{34})^{-\frac{k-1}{k(k+2)}}
x^{\frac{k-4}{2k(k+2)}} (1-x)^{\frac{1}{2k}} \cF^{(0)}_2 (x) \nonu
\left<\sigma_1 (w_1)\sigma_1 (w_2)\sigma_{k-1} (w_3)\sigma_{k-1} 
(w_4)\right>^{(1)} &= 
-(-1)^{\frac{k-2}{k(k+2)}} (w_{12}w_{34})^{-\frac{k-1}{k(k+2)}} x^{\frac{k-
4}{2k(k+2)}}
(1-x)^{\frac{1}{2k}} C \cF^{(1)}_2 (x) \ , \nonumber
\end{align}
where the $\cF^{(p)}_i$ are now $k$ dependent
\begin{align}
\cF^{(0)}_1 (x) &= x^{-\frac{3}{2(k+2)}} (1-x)^{\frac{1}{2(k+2)}}
F(\frac{1}{k+2},-\frac{1}{k+2},\frac{k}{k+2};x) \nonu
\cF^{(0)}_2 (x) &= \frac{1}{k} x^{\frac{2k+1}{2(k+2)}} (1-x)^{\frac{1}{2(k+2)}}
F(\frac{k+3}{k+2},\frac{k+1}{k+2},\frac{2k+2}{k+2};x) \nonu
\cF^{(1)}_1 (x) &= x^{\frac{1}{2(k+2)}} (1-x)^{\frac{1}{2(k+2)}}
F(\frac{1}{k+2},\frac{3}{k+2},\frac{k+4}{k+2};x) \nonu
\cF^{(1)}_2 (x) &= -2 x^{\frac{1}{2(k+2)}} (1-x)^{\frac{1}{2(k+2)}}
F(\frac{1}{k+2},\frac{3}{k+2},\frac{2}{k+2};x) \ .
\end{align}
{}From this, we find the following OPE coefficients
\begin{equation}
C_{\sigma_1,\sigma_1}^{\psi_1} = C_{\sigma_1,\psi_{k-1}}^{\sigma_{k-1}} = 
\frac{1}{\sqrt{k}} \ . 
\end{equation}

The wavefunction of four quasi-holes and $N=(r+1)k-2$ electrons
(with $r$ a positive integer) can be written in the
following form
\begin{multline}
\Psi^{(0,1)} (w_1,w_2,w_3,w_4;z_1,\ldots,z_N) =
\left<
\sigma_1 (w_1)\sigma_1 (w_2)\sigma_1 (w_3)\sigma_1 (w_4)
\psi_1 (z_1)\cdots \psi_1 (z_N)
\right>^{(0,1)} \times \\ \times
(w_{12}w_{34})^{\frac{3}{k(kM+2)}} x^{-\frac{2}{k(kM+2)}} (1-
x)^{\frac{1}{k(kM+2)}}
\prod_{i,j}(w_i-z_j)^{\frac{1}{k}}
\prod_{i<j}(z_i-z_j)^{\frac{kM+2}{k}} \ .
\label{rrkm}
\end{multline}
Again, we have the following master formula
\bea
\label{meqk}
\lefteqn{\Psi^{(0,1)} (w_1,w_2,w_3,w_4;z_1,\ldots,z_N) =} 
\nonu &&
A^{(0,1)} (\{w\}) \Psi_{(12)(34)} (\{w\},\{z\}) 
+ B^{(0,1)} (\{w\}) \Psi_{(13)(24)} (\{w\},\{z\}) \ .
\eea
To specify the functions
$\Psi_{(12)(34)}$ and $\Psi_{(13)(24)}$, we  
divide the electrons into $k$ groups, namely
$S_i = \{i,i+k,\ldots, N-(k-2)+i \}$ for $i=1,\ldots, k-2$ and 
$S_j = \{j,j+k,\ldots, N-(2k-2)+j \}$ for $j=k-1,k$. 
Setting $M=0$ for simplicity, we have
\begin{align}
\Psi_{(12)(34)} &= \frac{1}{\cN} \sum_{\{S_i\}} \Bigl[
\prod_{j \in S_{k-1}} (z_j-w_1)(z_j-w_2)
\prod_{j' \in S_{k}} (z_{j'}-w_3)(z_{j'}-w_4)
\prod_{i=1}^k \Psi^2_{S_i}
\Bigr] \nonu
\Psi_{(13)(24)} &= \frac{1}{\cN} \sum_{\{S_i\}} \Bigl[
\prod_{j \in S_{k-1}} (z_j-w_1)(z_j-w_3)
\prod_{j' \in S_{k}} (z_{j'}-w_2)(z_{j'}-w_4)
\prod_{i=1}^k \Psi^2_{S_i}
\Bigr] \ ,
\end{align}
where the sum is over all in-equivalent ways of dividing the
electrons in $k$ groups and $\cN = k^{\frac{N}{2}}/(k-2)!$.

We will consider the following two limits in the case of $N=2k-2$
electrons
\begin{equation}
(I) \quad 
\left\{ \begin{array}{l} 
z_i \to z_1 \qquad i=2,\ldots,k-1\\
z_1 \to w_2 \\
z_j \to z_{k} \qquad j = k+1,\ldots, 2 k-2\\ 
z_{k} \to w_4
\end{array}\right.
\quad
(II) \quad 
\left\{ \begin{array}{l}
z_i \to z_1 \qquad i=2,\ldots,k-1\\
z_1 \to w_3 \\
z_j \to z_{k} \qquad j = k+1,\ldots, 2 k-2\\ 
z_{k} \to w_4 \ .
\end{array} \right. 
\end{equation}
On the one hand, the master formula reduces to a form containing the
following two correlators
$\left< \sigma_1 \sigma_{k-1} \sigma_1 \sigma_{k-1}\right>$
and
$\left< \sigma_1 \sigma_1 \sigma_{k-1} \sigma_{k-1}\right>$
in the limits $(I)$ and $(II)$ respectively. On the other
hand, we find
\begin{align}
\lim_{(I)} \Psi_{(12)(34)} &= 
- \frac{((k-1)!)^2}{k^{k-1}} w_{14}w_{32}w_{42}^{2k-2} &
\lim_{(II)} \Psi_{(12)(34)} &= 0 \\
\lim_{(I)} \Psi_{(13)(24)} &= 0 &
\lim_{(II)} \Psi_{(13)(24)} &= 
\frac{((k-1)!)^2}{k^{k-1}} w_{14}w_{32}w_{34}^{2k-2} \ .
\nonumber
\end{align}

The functions $A^{(p)}$ and $B^{(p)}$ again follow
\begin{align}
A^{(0)} &= (w_{12}w_{34})^\alpha 
x^\beta (1-x)^\gamma \cF^{(0)}_1 (x) \nonu
B^{(0)} &= - (w_{12}w_{34})^\alpha
x^{\beta -1} (1-x)^{1+\gamma} \cF^{(0)}_2 (x) \nonu
A^{(1)} &= -(-1)^{\frac{2}{k+2}} C
(w_{12}w_{34})^\alpha x^\beta (1-x)^\gamma \cF^{(1)}_1 (x) \nonu
B^{(1)} &= (-1)^{\frac{2}{k+2}} C (w_{12}w_{34})^\alpha
x^{\beta -1} (1-x)^{1+\gamma} 
\cF^{(1)}_2 (x) \ , 
\end{align}
where we introduced the following notation
\begin{align}
\alpha & = \frac{2k+1}{2(k+2)}-\frac{3M}{2(kM+2)} &
\beta & = \frac{3}{2(k+2)}+\frac{2M}{2(kM+2)} & 
\gamma & = -\frac{M}{2(kM+2)} \ .
\end{align}

In this derivation, we used that 
the OPE coefficients for the parafermion fields $\psi_l$ are given by
\cite{fz1}
\begin{equation}
C_{\psi_l,\psi_{l'}}^2 = \frac{\Gamma(l+l'+1)\Gamma(k-l+1)\Gamma(k-l'+1)}
{\Gamma(l+1)\Gamma(l'+1)\Gamma(k-l-l'+1)\Gamma(k+1)} \ ,
\end{equation}
from which it follows that
\begin{equation}
\prod_{l=1}^{k-2} C_{\psi_1,\psi_l}^2 = \frac{((k-1)!)^2}{k^{k-2}} \ .
\end{equation}
In addition, we find the following braid matrices
\begin{align}
U_{12} &= (-1)^{-\frac{M}{2(kM+2)}}\begin{pmatrix}
(-1)^{\frac{2k+1}{2(k+2)}} & 0 \\ 0 & (-1)^{\frac{1}{2(k+2)}}
\end{pmatrix} 
\nonu
U_{23} &= \frac{(-1)^{-\frac{3}{2(k+2)}-\frac{3M}{2(kM+2)}}}{d_k} 
\begin{pmatrix}
1 & (-1)^{-\frac{2}{k+2}}\sqrt{d_k^2-1} \\ 
(-1)^{\frac{2}{k+2}}\sqrt{d_k^2-1} & -1
\end{pmatrix} 
\nonu
U_{13} &= \frac{(-1)^{\frac{3}{2(k+2)}-\frac{M}{2(kM+2)}}}{d_k} 
\begin{pmatrix}
1 & (-1)^{\frac{k-2}{k+2}}\sqrt{d_k^2-1} \\ 
-\sqrt{d_k^2-1} & (-1)^{\frac{k-2}{k+2}}
\end{pmatrix} \ ,
\label{rrgenbraid}
\end{align}
where $d_k = (C^0_0)^{-1}$.

\section{Detailed structure of the $su(3)_2$ parafermion theory}
\label{su3app}

We now turn the $su(3)_2$ parafermions, as introduced by Gepner in
\cite{gep1}.
This theory arises upon factoring two free fields from an
$SU(3)_2$ WZW theory, namely, it is the $su(3)_2/[u(1)]^2$ coset
CFT. In this appendix we present fusion rules, OPE's, 
4-point correlation functions and braiding properties.

\subsection{Fusion rules}
\label{sec:su32pf}

The $su(3)_2$ parafermion theory has 8 chiral sectors, which we label as 
$$\{ \id, \psi_1, \psi_2, \psi_{12}, \sigma_\up, \sigma_\down, 
\sigma_3, \rho\}\ .$$ 
It is important to realize that each of these sectors contains an infinite
number of Virasoro primary fields. We sometimes use the same notation for
the parafermion sector (say $\psi_1$) and for the leading Virasoro primary 
field ($\psi_1(z)$). We shall see that in the sector denoted as $\rho$
there are two independent leading Virasoro primaries, which we shall write as
$\rho_s(z)$ and $\rho_c(z)$. 

At the level of parafermion sectors, the merging of two sectors is expressed 
via {\it fusion rules}, which we present in this section. Our computations in
section \ref{sec:asstates} require more detailed information at the level of 
(primary) fields.
The latter is contained in the OPE's that we present in the next section
of this appendix. 

The fusion rules of the eight parafermion sectors can be derived from
the coset description of this theory. For our purposes, we will use the
defining coset $su(3)_2/[u(1)]^2$, as given in \cite{gep1}.
The parafermion sectors are labeled by an $su(3)$ label
$\Lambda=(\Lambda_1,\Lambda_2)$ (when labeling the representations
in terms of the dimensions, we would have $\id = (0,0)$, ${\bf 3} = (1,0)$,
${\bf \bar{3}} = (0,1)$, ${\bf 6} = (2,0)$, ${\bf \bar{6}} = (0,2)$ and ${\bf 8} = (1,1)$)
and two $u(1)$ labels $\lambda=(\lambda_1,\lambda_2)$.

There are various restrictions on the labels. First of all, we have the
branching condition
$\Lambda_1+2\Lambda_2 =(\lambda_1+2\lambda_2) \mod 3$.
The label $\lambda$ is only defined up to $2$ times (in general, $k$ times)
the root lattice of $su(3)$, which means the following sectors are identified
\begin{align}
\label{u1ident}
\Phi^\Lambda_{\lambda_1,\lambda_2}&\equiv
\Phi^\Lambda_{\lambda_1+4,\lambda_2-2} & 
\Phi^\Lambda_{\lambda_1,\lambda_2}&\equiv
\Phi^\Lambda_{\lambda_1-2,\lambda_2+4} \ .
\end{align}
In addition, there are other identifications, which follow from
the structure of the affine Lie algebra $su(3)_2$, see \cite{gep2}
\begin{equation}
\label{fieldident}
\Phi^{(\Lambda_1,\Lambda_2)}_{(\lambda_1,\lambda_2)} \equiv
\Phi^{(2-\Lambda_1-\Lambda_2,\Lambda_1)}_{(\lambda_1+2,\lambda_2)} \equiv
\Phi^{(\Lambda_2,2-\Lambda_1-\Lambda_2)}_{(\lambda_1,\lambda_2+2)} \ .
\end{equation}

{}From these rules, it follows that there are indeed eight different 
sectors, or `parafermion fields', as mentioned above. The fusion rules 
follow from the general rule
\begin{equation}
\Phi^\Lambda_\lambda \times \Phi^{\Lambda'}_{\lambda'} = 
\sum_{\Lambda'' \in \Lambda \times \Lambda'}
\Phi^{\Lambda''}_{\lambda+\lambda'} \ .
\end{equation}

The fusion rules can now easily be derived from the following set
of `representations' of the eight parafermion sectors, which close under
fusion
\begin{align}
\id & = \Phi^{(0,0)}_{(0,0)} &
\psi_1 & = \Phi^{(0,0)}_{(2,-1)} &
\psi_2 & = \Phi^{(0,0)}_{(1,-2)} &
\psi_{12} & = \Phi^{(0,0)}_{(1,1)}  
\nonu
\sigma_\uparrow & =  \Phi^{(1,1)}_{(-1,2)} &
\sigma_\downarrow & = \Phi^{(1,1)}_{(2,-1)} &
\sigma_3 & = \Phi^{(1,1)}_{(1,1)} &
\rho & = \Phi^{(1,1)}_{(0,0)} \ .
\label{su32pf}
\end{align}
Using the only non-trivial fusion rule of the $su(3)_2$ fields 
$(0,0)$ and $(1,1)$
(corresponding to the one and eight dimensional representation, respectively), 
namely,
$(1,1)\times(1,1)=(0,0)+(1,1)$ and the identifications \eqref{u1ident},
we find the fusion rules as given in table \ref{fusrul}. 

\begin{table}[ht]
\begin{center}
\begin{tabular}{|c|c|c|c|c|c|c|c|}
\hline
$\times$ & $\sigma_\uparrow$ & $\sigma_\downarrow$ &
$\sigma_3$ & $\rho$ & $\psi_1$ & $\psi_2$ & $\psi_{12}$
\\ \hline
$\sigma_\uparrow$ & $\id + \rho$ &&&&&&  \\

$\sigma_\downarrow$ & $\psi_{12}+\sigma_3$ & $\id+\rho$ &&&&&\\

$\sigma_3$ & $\psi_1 + \sigma_\downarrow$ &
$\psi_2 + \sigma_\uparrow$ & $\id + \rho$ &&&&\\

$\rho$ & $\psi_2$ + $\sigma_\uparrow$ &
$\psi_1 + \sigma_\downarrow$ & $\psi_{12} + \sigma_3$ &
$\id + \rho$ &&&\\

$\psi_1$ & $\sigma_3$ & $\rho$ & $\sigma_\uparrow$ &
$\sigma_\downarrow$ & $\id$ &&\\

$\psi_2$ & $\rho$ & $\sigma_3$ & $\sigma_\downarrow$ &
$\sigma_\uparrow$ & $\psi_{12}$ & $\id$ &\\

$\psi_{12} $ & $ \sigma_\downarrow$ & $\sigma_\uparrow$ &
$\rho$ & $\sigma_3$ & $\psi_2$ & $\psi_1$ & $\id$ \\
\hline
\end{tabular}
\end{center}
\vskip 2mm \caption{Fusion rules of the parafermion and spin
sectors associated to the parafermion theory $su(3)_2/[u(1)]^2$
introduced by Gepner \protect\cite{gep1}.} \label{fusrul}
\end{table}
Of course, these fusion rules can also be derived from the
$S$-matrix and the Verlinde formula. 

\subsection{OPE's}
\label{opesu32}

To our knowledge, the OPE's of the leading Virasoro primary fields in
the various sectors of the $su(3)_2$ parafermion theory have not been 
presented in the literature.

We have determined the leading terms in the OPE's of these Virasoro 
primaries. We observed a $\ZZ_3$ symmetry relating $(\sd,\su,\sthree)$ 
and $(\psi_1,\psi_2,\sqrt{2}\psi_{12})$. To streamline notations, we 
therefore write $\sigma_1=\sigma_\down$, $\sigma_2=\sigma_\up$, and 
$\psi_3=\sqrt{2}\psi_{12}$. We also employ Virasoro primaries
$\rho_i$ and $\bar{\rho}_i$, $i=1,2,3$. These fields are all linear
combinations (specified below) of the fields $\rho_s$ and $\rho_c$.
The scaling dimensions of the leading fields are
\be
h_{\psi}=\frac{1}{2}\ , \quad
h_{\sigma}=\frac{1}{10}\ , \quad
h_{\rho}=\frac{3}{5}\ .
\ee

For fixing the details of the OPE's, especially those involving the
fields $\rho_s$, $\rho_c$, we have relied on various contractions of 
the master formulas such as eq.~(\ref{eq:master3333}), 
(\ref{eq:masterupupdowndown}), etc., presented in section \ref{sec:qhsu32}. 
These formulas provide expressions for correlators of four spin
fields $\sigma_{i}$ plus an arbitrary number of parafermions
$\psi_{j}$. By fusing some of the $\sigma_{i}$ with the 
$\psi_{j}$, one produces various combinations of the fields
$\rho_{s,c}$\ ; in the end this gives enough information to uniquely
fix the OPE's. [We remark that the logical structure of our 
reasoning has been quite delicate: we have relied on the general `master
formula' structure of correlators and on the `seed' provided by the 
explicit correlators provided by the KZ paper and by some of the 
simplest OPE's, set by the parafermion fusion rules. Combined these
turn out to be strong enough to fix both the coefficients in the master 
formulas and the OPE's of all fields involved.]  In section 
\ref{sec:morecorrsu32} below we explicitly mention some of the contractions 
we used and we provide some additional correlation functions of the 
parafermion theory.

Employing the same schematic notation as in section \ref{aope},
eq.\ \eqref{eq:OPEAB1}, \eqref{eq:OPEAB2}, we have the following OPE's 
\bea
\psi_i \psi_j &=& \id \qquad {\rm for} \ i=j 
\nonumber \\
&=& \frac{1}{\sqrt{2}} \psi_k \qquad {\rm for} 
     \ i\neq j, \ i\neq k, \ j\neq k
\nonu
\psi_i \sigma_j &=& \bar{\rho}_i 
                    \qquad {\rm for} \ i=j
\nonumber \\
&=& \frac{1}{\sqrt{2}} \sigma_k \qquad {\rm for} 
     \ i\neq j, \ i\neq k, \ j\neq k
\nonu
\sigma_i \sigma_j &=& \id + \sqrt{2C} \rho_i \qquad {\rm for} \ i=j 
\nonumber \\
&=& \frac{1}{\sqrt{2}} \psi_k + \sqrt{-3C} \ \sigma_k 
    \qquad {\rm for} \ i\neq j, \ i\neq k, \ j\neq k
\nonu
\bar{\rho}_i \psi_j &=& \sigma_j \qquad {\rm for} \ i=j 
\nonumber \\
                    &=& - \frac{1}{2} \sigma_j \qquad {\rm for} \ i\neq j
\nonu
 (\bar{\rho}_i \sigma_j)^{(0)} &= 
 & \psi_j + \ldots \qquad {\rm for} \ i=j 
\nonumber \\
                    & = 
                    & - \frac{1}{2} \psi_j 
                        + \ldots \qquad {\rm for} \ i\neq j
\nonu
(\rho_i \sigma_j)^{(1)} &= 
 & \sqrt{2C} \sigma_j + \ldots \qquad {\rm for} \ i=j 
\nonumber \\
            & = 
            & - \frac{1}{2} \sqrt{2C} \sigma_j 
                + \ldots \qquad {\rm for} \ i\neq j
\eea
Note that we have written separate equations for fusing
fields from the $\rho$ and $\sigma$ sectors into the $(0)$
channel or the $(1)$ channel.

The $\bar{\rho}_i$ and $\rho_i$ can all be expressed in two
independent Virasoro primaries $\rho_s$ and $\rho_c$ according 
to
\bea
&&\rho_1= \frac{1}{2} (\rho_c - \sqrt{3} \rho_s) \ , \quad
  \rho_2= \frac{1}{2} (\rho_c + \sqrt{3} \rho_s) \ , \quad
  \rho_3= - \rho_c \ ,
\nonu
&& 
\bar{\rho}_1= \frac{1}{2} (\sqrt{3} \rho_c +\rho_s) \ , \quad
\bar{\rho}_2= \frac{1}{2} (- \sqrt{3} \rho_c + \rho_s) \ , \quad
\bar{\rho}_3=  - \rho_s \ ,
\label{eq:rhos}
\eea
The list of basic OPE's is then completed by
\bea
\rho_s \rho_s = \id - \sqrt{2C} \rho_c \ , & \quad
\rho_s \rho_c = - \sqrt{2C} \rho_s \ , & \quad
\rho_c \rho_c = \id + \sqrt{2C} \rho_c 
\label{eq:rhorhorho}
\eea
giving
\bea
\rho_i \rho_j &=& \id - \sqrt{2C} \rho_i \qquad {\rm for} \ i=j
\nonumber\\              
              &=& - \frac{1}{2} \id - \sqrt{2C} \rho_k \qquad {\rm for} 
                  \ i\neq j, \ i\neq k, \ j\neq k
\nonu
\bar{\rho}_i \bar{\rho}_j &=& \id + \sqrt{2C} \rho_i \qquad {\rm for} \ i=j
\nonumber\\              &=& -\frac{1}{2} \id 
                  + \sqrt{2C} \rho_k \qquad {\rm for} 
                  \ i\neq j, \ i\neq k, \ j\neq k
\eea
In all these expressions, the constant $C$ has the value 
given below eq.\ (\ref{eq:four3w}).

\subsection{Spin field correlators}

We now present the four-point correlation functions for the spin fields 
in the parafermion theory. They are obtained by factoring the four point 
functions of fundamental fields in the $SU(3)_2$ WZW model, as given by
\cite{kz1}, by factors associated to the spin and charge bosons
\begin{align}
\label{eq:fourspin}
\langle \su(w_1)\su(w_2)\sd(w_3)\sd(w_4)\rangle^{(0)} &=
(w_{12}w_{34})^{-\frac{1}{5}} x^{\frac{8}{15}}
(1-x)^{-\frac{1}{6}} {\cal F}_1^{(0)}(x) 
\\[2mm]
\langle \su(w_1)\su(w_2)\sd(w_3)\sd(w_4)\rangle^{(1)} &= 
-(-1)^{\frac{3}{5}}C (w_{12}w_{34})^{-\frac{1}{5}} x^{\frac{8}{15}}
(1-x)^{-\frac{1}{6}} {\cal F}_1^{(1)}(x) 
\nonu
\langle \su(w_1)\sd(w_2)\su(w_3)\sd(w_4)\rangle^{(0)} &=
(-1)^{\frac{1}{2}} (w_{12}w_{34})^{-\frac{1}{5}} x^{\frac{1}{30}}
(1-x)^{\frac{1}{3}} {\cal F}_2^{(0)}(x) 
\nonu
\langle \su(w_1)\sd(w_2)\su(w_3)\sd(w_4)\rangle^{(1)} &=
(-1)^{\frac{1}{10}}C (w_{12}w_{34})^{-\frac{1}{5}} x^{\frac{1}{30}}
(1-x)^{\frac{1}{3}} {\cal F}_2^{(1)}(x) 
\nonu
\langle \sigma_3(w_1)\sigma_3(w_2)\sigma_3(w_3)\sigma_3(w_4)\rangle^{(0)} &=
(w_{12}w_{34})^{-\frac{1}{5}} x^{\frac{8}{15}}
(1-x)^{\frac{1}{3}}[{\cal F}_1^{(0)}(x)+{\cal F}_2^{(0)}(x)] 
\nonu
\langle \sigma_3(w_1)\sigma_3(w_2)\sigma_3(w_3)\sigma_3(w_4)\rangle^{(1)} &=
-(-1)^{\frac{3}{5}}C (w_{12}w_{34})^{-\frac{1}{5}} x^{\frac{8}{15}}
(1-x)^{\frac{1}{3}} [{\cal F}_1^{(1)}(x) + {\cal F}_2^{(1)}(x)] \ , \nonumber
\end{align}
where $x$ is as usual $x=\frac{(w_1-w_2)(w_3-w_4)}{(w_1-w_4)(w_3-w_2)}$.
The functions ${\cal F}_{i}^{(p)}(x)$ are now given by
\begin{align}
\cF^{(0)}_1 (x) &= x^{-\frac{8}{15}} (1-x)^{\frac{1}{15}}
F(\frac{1}{5},-\frac{1}{5},\frac{2}{5};x)  &
\cF^{(0)}_2 (x) &= \frac{1}{2} x^{\frac{7}{15}} (1-x)^{\frac{1}{15}}
F(\frac{6}{5},\frac{4}{5},\frac{7}{5};x) 
\nonu
\cF^{(1)}_1 (x) &= x^{\frac{1}{15}} (1-x)^{\frac{1}{15}}
F(\frac{2}{5},\frac{4}{5},\frac{8}{5};x)  &
\cF^{(1)}_2 (x) &= -3 x^{\frac{1}{15}} (1-x)^{\frac{1}{15}}
F(\frac{2}{5},\frac{4}{5},\frac{3}{5};x) \ .
\label{fsu32}
\end{align}

\subsection{Further correlators}
\label{sec:morecorrsu32}

The master formulas developed in section \ref{sec:qhsu32} can
be used to gain insight into the OPE's and correlation functions
involving the fields in the $\rho$ sector of the $su(3)_2$ parafermion
theory. In section \ref{opesu32} we already displayed some of the
OPE's satisfied by $\rho_s$ and $\rho_c$.  In deriving these we proceeded
as follows. Having defined the combinations $\rho_i$ and $\bar{\rho}_i$
through the OPE's $(\sigma_i \sigma_i)^{(1)} = \sqrt{2C} \rho_i$ and
$(\psi_i \sigma_i)^{(1)} = \bar{\rho}_i$, we observed that a specific 
contraction of the master formula eq.~(\ref{eq:masterupupdowndown}) 
shows that $\langle \sigma_i \sigma_i \bar{\rho}_i \rangle$ vanishes.
This implies that $\rho_i$ and $\bar{\rho}_i$ are orthogonal (have a 
vanishing 2-point function). Exploiting the symmetry among the
$i=1,2,3$ labels leads to the parametrization eq.~(\ref{eq:rhos}).
All this leaves some freedom in the self-couplings of the $\rho_c$
and $\rho_s$. To fix these, we observed that yet another contraction of 
the master formula implies that the three point function
$\langle \bar{\rho}_1 \bar{\rho}_2 \bar{\rho}_3 \rangle$ vanishes.
This contraction arises from the formula 
\begin{eqnarray}
\lefteqn{
\langle
\sigma_1(w_1)\sigma_2(w_2)\sigma_3(w_3)
\psi_1(z_1)\psi_2(z_2)\psi_3(z_3)
\rangle =}
\nonu
&& (-1)^{-\frac{9}{10}} \frac{1}{2} \sqrt{\frac{3C}{2}}
     ( z_{12} z_{23} z_{31})^{-\frac{1}{2}}
     ( w_{12} w_{23} w_{31})^{-\frac{1}{10}}
\nonu
&& \times
     \left[ 
     \frac{(z_1-w_2)(z_2-w_3)(z_3-w_1)+(z_1-w_3)(z_2-w_1)(z_3-w_2)}
     {[(z_1-w_2)(z_2-w_3)(z_3-w_1)(z_1-w_3)(z_2-w_1)(z_3-w_2)]^{\frac{1}{2}}}
     \right]
\end{eqnarray}
in the limit where $z_i \to w_i$ for $i=1,2,3$.

The vanishing of 
$\langle \bar{\rho}_1 \bar{\rho}_2 \bar{\rho}_3 \rangle$
implies that the combination of $\rho_c$ and $\rho_s$ featuring
in the $(1)$ channel of the fusion product of $\bar{\rho}_1$ and 
$\bar{\rho}_2$ is orthogonal to $\bar{\rho}_3$ and thereby proportional
to $\rho_3$. A final contraction yielding $\langle \sigma_2 \sigma_2 
\bar{\rho}_1 \bar{\rho}_1 \rangle$ is then used to fix the normalization, 
giving eq.~(\ref{eq:rhorhorho}).

Correlation functions involving one or more fields in the $\rho$
sector are easily generated as suitable contractions of the various
master formulas. Examples are the following four point functions
\bea
\label{eq:fourrho}
\lefteqn{\langle \bar{\rho}_1(w_1) \bar{\rho}_1(w_2)
        \bar{\rho}_1(w_3) \bar{\rho}_1(w_4) \rangle^{(0)} =}
\nonu &&
(w_{12}w_{34})^{-\frac{6}{5}} x^{\frac{8}{15}}
(1-x)^{-\frac{2}{3}} 
(1-x+x^2)\left[ {\cal F}_1^{(0)}(x) + {\cal F}_2^{(0)}(x) \right] 
\nonu
\lefteqn{\langle \bar{\rho}_1(w_1) \bar{\rho}_1(w_2)
        \bar{\rho}_1(w_3) \bar{\rho}_1(w_4) \rangle^{(1)} =}
\nonu &&
-(-1)^{\frac{3}{5}}C (w_{12}w_{34})^{-\frac{6}{5}} x^{\frac{8}{15}}
(1-x)^{-\frac{2}{3}} 
(1-x+x^2)\left[ {\cal F}_1^{(1)}(x) + {\cal F}_2^{(1)}(x) \right] 
\nonu
\lefteqn{\langle \bar{\rho}_1(w_1) \bar{\rho}_1(w_2)
        \bar{\rho}_2(w_3) \bar{\rho}_2(w_4) \rangle^{(0)} =}
\nonu &&
\frac{1}{4} (w_{12}w_{34})^{-\frac{6}{5}} x^{\frac{8}{15}}
(1-x)^{\frac{4}{3}} 
\left[ \left(1+\frac{3}{(1-x)^2}\right) {\cal F}_1^{(0)}(x) 
      + \frac{x^2}{(1-x)^2} {\cal F}_2^{(0)}(x) \right] 
\\[2mm]
\lefteqn{\langle \bar{\rho}_1(w_1) \bar{\rho}_1(w_2)
        \bar{\rho}_2(w_3) \bar{\rho}_2(w_4) \rangle^{(1)} =}
\nonu &&
-(-1)^\frac{3}{5} \frac{C}{4} (w_{12}w_{34})^{-\frac{6}{5}} 
x^{\frac{8}{15}}(1-x)^{\frac{4}{3}} 
\left[ \left(1+\frac{3}{(1-x)^2}\right) {\cal F}_1^{(1)}(x) 
      + \frac{x^2}{(1-x)^2} {\cal F}_2^{(1)}(x) \right] \ .
\nonumber
\eea

\subsection{Braiding relations}
\label{su32braid}

We list the transformation properties of the functions 
${\cal F}_{1,2}^{(0,1)}(x)$, as specified in eq.\ (\ref{fsu32})
under transformations (i) $x \to 1-x$, (ii) $x \to \frac{-x}{1-x}$, 
(iii) $x \to \frac{1}{x}$.

For $x \to 1-x$ we have \cite{kz1}
\be
{\cal F}_{1,2}^{(p)}(1-x) = \sum_q C^p_q{\cal F}_{2,1}^{(q)}(x) 
\ee
with
\bea
&&
C_0^0=-C_1^1=3\frac{\Gamma(\frac{3}{5})\Gamma(-\frac{3}{5})}
    {\Gamma(\frac{1}{5})\Gamma(-\frac{1}{5})} 
    = \frac{1}{2}(\sqrt{5} - 1)= \tau \ .
\nonu
&& 
C_0^1=-3\frac{ \Gamma^2(\frac{3}{5})}
    {\Gamma(\frac{2}{5})\Gamma(\frac{4}{5})}  \ , \qquad
C_1^0 = \frac{1+ C_0^0 C_1^1}{C_0^1}
     = - \frac{1}{3} \frac{\Gamma^2(\frac{2}{5})}
    {\Gamma(\frac{1}{5})\Gamma(\frac{3}{5})} \, .
\eea
We observe that
\be
C^2=C_1^0/C_0^1=
\frac{1}{9} \frac{\Gamma(\frac{4}{5})\Gamma^3(\frac{2}{5})}
                 {\Gamma(\frac{1}{5})\Gamma^3 (\frac{3}{5})}
 \ , \qquad
C_1^0/C = C_0^1 C = - \sqrt{\tau}  \  .
\ee
For $x \to \frac{-x}{1-x}$ we have
\bea
&&
{\cal F}_1^{(0)}(\frac{-x}{1-x}) = 
         (-1)^{-\frac{8}{15}} (1-x)^{\frac{1}{5}} {\cal F}_1^{(0)}(x) 
\nonu
&&
{\cal F}_2^{(0)}(\frac{-x}{1-x}) = 
         (-1)^{-\frac{8}{15}} (1-x)^{\frac{1}{5}} 
         [-x {\cal F}_1^{(0)}(x)+(1-x) {\cal F}_2^{(0)}(x)]
\nonu
&&
{\cal F}_1^{(1)}(\frac{-x}{1-x}) =
         (-1)^{\frac{1}{15}} (1-x)^{\frac{1}{5}} 
         {\cal F}_1^{(1)}(x) 
\nonu
&&
{\cal F}_2^{(1)}(\frac{-x}{1-x}) = 
         (-1)^{\frac{1}{15}} (1-x)^{\frac{1}{5}} 
         [-x {\cal F}_1^{(1)}(x)+(1-x) {\cal F}_2^{(1)}(x)] \, .
\eea
Finally for $x \to \frac{1}{x}$ 
\bea
&&
{\cal F}_2^{(0)}(\frac{1}{x}) = 
       (-1)^{-\frac{2}{15}} x^{\frac{1}{5}} 
       [C_0^0 {\cal F}_2^{(0)}(x) - (-1)^{\frac{2}{5}} C_1^0 {\cal F}_2^{(1)}(x)]
\nonu
&&
{\cal F}_2^{(1)}(\frac{1}{x}) = 
       (-1)^{-\frac{2}{15}} x^{\frac{1}{5}} 
       [- (-1)^{\frac{2}{5}} C_0^1 {\cal F}_2^{(0)}(x) 
        - (-1)^{-1/5} C_1^1 {\cal F}_2^{(1)}(x))]
\nonu
&&
{\cal F}_1^{(0)}(\frac{1}{x}) = 
       (-1)^{-\frac{2}{15}} x^{\frac{1}{5}} 
       [C_0^0 \, \bigl( x {\cal F}_1^{(0)}(x) - (1-x){\cal F}_2^{(0)}(x) \bigr) 
\nonu
&& \qquad \qquad \qquad \qquad
        - (-1)^{\frac{2}{5}} C_1^0 \, 
               \bigl( x {\cal F}_1^{(1)}(x) - (1-x){\cal F}_2^{(1)}(x) \bigr) ]
\nonu
&&
{\cal F}_1^{(1)}(\frac{1}{x}) = 
       (-1)^{-\frac{2}{15}} x^{\frac{1}{5}} 
       [- (-1)^{\frac{2}{5}} C_0^1 \, \bigl(x 
        {\cal F}_1^{(0)}(x) - (1-x){\cal F}_2^{(0)}(x) \bigr) 
\nonu
&& \qquad \qquad \qquad \qquad
        - (-1)^{-\frac{1}{5}} C_1^1 \,
        \bigl( x {\cal F}_1^{(1)}(x) - (1-x){\cal F}_2^{(1)}(x) \bigr) ] \, .
\eea

\end{document}